\documentclass[preprint]{aastex}
\usepackage{natbib}
\usepackage{amssymb,amsmath,fullpage,epsfig,graphicx,bm}
\shorttitle{Towards Quark Deconfinement via spindown of Neutron Stars}
\shortauthors{Staff et al.}

\begin{document}

\title{Spindown of Isolated Neutron Stars: \\
Gravitational Waves or Magnetic Braking?}

\author{Jan E. Staff}
\affil{Department of Physics and Astronomy, Louisiana State University,
202 Nicholson Hall, Tower Dr., Baton Rouge, LA 70803-4001, 
USA}
\author{Prashanth Jaikumar, Vincent Chan}
\affil{Department of Physics \& Astronomy, California State University Long Beach,\\ 1250 Bellflower Blvd., Long Beach, California 90840, USA}
\author{Rachid Ouyed}
\affil{Department of Physics and Astronomy, University of Calgary,
2500 University Drive NW, Calgary, Alberta, T2N 1N4 Canada}

\begin{abstract}  

We study the spindown of isolated neutron stars from initially rapid
rotation rates, driven by two factors: (i) gravitational wave emission due
to r-modes and (ii) magnetic braking.  In the context of isolated neutron
stars, we present the first study including self-consistently the
magnetic damping of r-modes in the spin evolution.  We track the spin
evolution employing the RNS code, which accounts for the rotating structure
of neutron stars for various equations of state.  We find that, despite the
strong damping due to the magnetic field, r-modes alter the braking rate from pure magnetic braking for $B\leq 10^{13}$G. For realistic values of the saturation amplitude $\alpha_{\rm sat}$, the r-mode can also decrease the
time to reach the threshold central density for quark deconfinement. Within a phenomenological model, we assess the gravitational waveform that would result from r-mode driven spindown of a magnetized neutron star.  To contrast with the persistent signal during the spindown phase, we also present a preliminary estimate of the transient gravitational wave signal from an explosive quark-hadron phase transition, which can be a signal for the deconfinement of quarks inside neutron stars.

\newpage
\end{abstract}

\section{Introduction}

Neutron stars are highly compact stars of typical radius $R\sim 12$ km and mass $M\sim 1.5M_{\odot}$ made mostly of degenerate neutron-rich matter 
at densities up to several times nuclear matter saturation density $\rho_0=2.5\times 10^{14}$ g/cc. By tracking their long-term thermal and
rotational evolution, we can learn about the nature of matter under the
crust. For example, \citet{Page11} have proposed that the thermal history of the neutron star in Cas A may be indicating the recent onset of neutron superfluidity deep in its interior. Recently, \citet{Neg11} have shown how rotational evolution is linked to a reorganization of particle composition in the stellar interior, leading to switching on of exotic neutrino emission processes. A neutron star is a complex system with intertwined physical properties that can change on relatively short astrophysical timescales. 

In this paper, we will be concerned with the rotational evolution of a
highly magnetized isolated neutron star.  The electromagnetic emissions of a
neutron star derive from its rotational kinetic energy, and its spindown is
usually measured in terms of a braking index, $n$, which is dependent on the
magnetic field configuration ($n$=3 for a dipolar field).  Neutron stars can
also spindown through gravitational wave emissions associated to the r-mode
\citep{Anders}.  In fact, the observational interest in r-modes comes from
the fact that no neutron stars have been found to spin at rates near the
maximum allowed frequency (the ``break-up frequency'').  r-modes offer a
possible explanation of this fact: in rotating neutron stars, these modes
lose energy through gravitational waves, which carry away angular momentum
from the star and act as braking radiation.  As the star spins down, its
central density increases.  This increase could be sufficient to make
baryonic matter undergo phase transitions to more exotic phases of strongly
interacting matter, such as quark matter, with possible implications for
gamma-ray bursts \citep{Bez,HZ,Drago2} and the formation of quark stars, if
theoretical conjectures about the absolute stability of strange quark matter
are realized in nature \citep{Itoh, Bodmer, Witten}. While it is almost
certain that this quark phase, if it exists inside neutron stars, cannot be
a free gas of quarks \citep{Ozel,Weissenborn11}, an interacting phase of
quarks still appears to be consistent with the recent finding of a
2$M_{\odot}$ neutron star \citep{Demorest}.

The main questions we seek to answer are: given a few initial
parameters of the newly born neutron star, such as its spin period, magnetic
field and baryonic mass, can we determine which neutron stars are likely to
eventually manifest a quark phase in their interior?  If so, how long before
the transition to quark matter occurs?  In this work, we take a step towards
answering these questions by taking a closer look at the rotational
evolution of a newly-born hot neutron star as it spins down, and we focus on two main driving factors - magnetic braking and gravitational radiation from r-modes. We consider different equations of state (EoS) for neutron stars,
since the EoS at high density is uncertain and lacking strong empirical
constraints \citep[although a recent statistical analysis shows that the equation
of state stiffens at high densities and is consistent with the expected
range in certain nuclear parameters;][]{Steiner10}.  Furthermore, second
generation gravitational wave detectors such as Advanced LIGO will soon be
operational, and our work provides an update for a similar theoretical study
\citep{Lai} performed several years ago with regard to the LIGO detector. 
We should note here that in contrast to the assumptions in \citet{Lai},
our work includes the effects of magnetic damping of r-modes in the spindown
evolution, which leads to quantitatively different results.

In section \ref{Theorymag}, we outline the approach and main equations that describe the neutron star's spindown. The theoretical analysis follows in part the works of \citet{Lai}, as well as \citet{Drago1} and
\citet{Rezzolla}. Section \ref{Magresults} collects our main conclusions from this analysis. In section \ref{GWevo}, we analyze the evolution of the gravitational wave frequency associated to the growing r-mode, along the lines of \citet{Owen98}, and present results in section \ref{GWresults}. We conclude in section \ref{conclude} with a preliminary calculation of the gravitational wave signal from an explosive deconfinement transition (the ``Quark-Nova").

\section{Spindown to deconfinement} \label{Theorymag} In a previous paper
\citep{Staff1}, we had determined birth parameters of a neutron star which
would support the transition to deconfinement driven solely by magnetic
braking.  We found that neutron stars that are born with mass $M\gtrsim
1.5M_{\odot}$ and spin-period $P\lesssim 3$ ms are the best candidates to
reach deconfinement density in their core (assumed to be $\sim 5\rho_0$),
given a range of neutron star magnetic fields $10^{12}$-$10^{15}$ Gauss. 
This small value of the initial spin-period (or large spin frequency) is
required since magnetic braking from a rotating magnetic dipole is itself
proportional to the third power of the spin frequency.  Most of the
increase in central density occurs soon after the rapidly rotating neutron
star is born.

In that work, we neglected the role of gravitational radiation from r-modes in an adhoc way - by assuming that neutrons stars are born axisymmetric and remain so. In reality, non-axisymmetric perturbations are expected due to the violent process of a supernova, which leaves the new-born neutron star in a turbulent state \citep{Keil,Akiyama}. What is then the role of r-modes and gravitational braking in comparison to magnetic braking as far as the time to deconfinement is concerned? Previous studies on the interplay of r-modes and neutron star magnetic fields have focused on different questions - for example, \citet{Lee} has studied the r-mode in magnetized neutron stars with $B\leq 10^{12}$G with an aim to understand X-ray pulsations from local hot spots in accretion-powered pulsars. \citet{Lai} have shown that magnetic fields of $B\geq 10^{14}$G can make magnetic braking as important as r-mode spindown, specially for slowly rotating stars, and that Alfv\'en wave driving of the r-mode can also play a role. 

\vskip 0.2cm

 Meanwhile, other works \citep{Shapiro,Rezzolla,Kiuchi,Drago1} have
focused on the evolution of toroidal magnetic fields generated by the
secular r-mode.  Essentially, the toroidal field is generated by
differential rotation of the stellar fluid associated with the r-mode.  The
differential rotation is a key feature of the r-mode instability in the
non-linear regime \citep{Levin}.  \citet{Shapiro} have shown that, for
isolated neutron stars, this effect can amplify existing magnetic fields by
two orders of magnitude within the time taken by the r-mode to saturate
($\approx$ few hundred secs.).  Back-reaction on the r-mode due to this
toroidal field implies that there is an associated magnetic damping, which
has to be factored into the spin evolution.  For neutron stars that are
accreting from a binary companion, the build-up of toroidal fields to
sizeable values is rather slow, and according to \citet{Drago1}, can take up
to several hundred years.  However, we are interested in this effect on
isolated neutron stars.  The magnitude of the magnetic damping term $F_m$ is
proportional to the integrated time evolution of $\alpha^2(t)$
\citep{Rezzolla}, with $\alpha$ being the r-mode amplitude. These authors
obtained a strong damping effect by assuming that the background evolution
of $\alpha(t)$ is identical to that given by the absence of any magnetic
fields.  However, we solve the evolution equations for the star's angular
frequency $\Omega$ and $\alpha(t)$ with magnetic fields and magnetic damping
from the start. $F_m$ grows while the mode is unstable, and once it saturates $F_m$ remains constant.  This consistent inclusion of $F_m$ leads to a slower evolutionary path for $\alpha(t)$ towards saturation, hence the magnetic damping is not as effective in suppressing the r-mode.  Consequently, for a range of magnetic fields and small r-mode saturation values, we find that this timescale can be much larger than the typical time taken by the neutron star to spindown
to typical quark deconfinement density.  Therefore, our results clearly
indicate that the r-mode, for realistic saturation values, quantitatively
affects the spindown evolution of an isolated neutron star. We now discuss
the relevant equations for the evolution of the star's rotation rate and
r-mode evolution.

The spindown $\dot{\Omega}$ of a neutron star (mass $M$, radius $R$) is
accompanied by a loss of energy as well as angular momentum.  For the
radiating star, conservation of total angular momentum $J_{\rm tot}=J_{\rm
star}+(1-K_j)J_c$ with $K_j$ a dimensionless constant and $J_{\rm
star}=I\Omega$ (where $I$ is the moment of inertia about the rotation axis)
yields \citep{Wagoner, Drago1}

\begin{equation}
\frac{dJ_{\rm tot}}{dt}=2J_cF_g+\dot{J_a}-I\Omega F_{\rm mag} \,.
\end{equation}

$J_c=-K_c\alpha^2 J_{\rm star}$ is the canonical angular momentum of the r-mode to 1st order in $\Omega$ \citep{Schutz}. $F_g$ is the rate of gravitational radiation associated to the $l$=$m$=2 current multipole taken for an $n$=1 polytropic star from eqn.(65) of \citet{Andy1}. This is expected to be a good approximation to a large part of the neutron star interior. $\dot{J_a}$ is the accretion rate onto the star (assumed zero for our case since we consider isolated stars only) and $F_{\rm mag}$ is the magnetic braking rate \citep{Manchester}. The dimensionless quantity $K_c=3\bar{J}/2\bar{I}$ is defined from 

\begin{eqnarray}
\bar{J}=\frac{1}{MR^4}\int_0^R \rho r^6 dr \,,\nonumber\\
\bar{I}=\frac{8\pi}{3MR^2}\int_0^R\rho r^4 dr\,.
\label{IandJbar}
\end{eqnarray}

where $\rho=\rho(r)$ is the density profile (assumed radially symmetric) of
the star~\footnote{ In principle, equatorial flattening due to rotation
renders the density profile asymmetric, but this does not change the value
of $K_c$ significantly, which is the basis for our approximation of a
symmetric profile.}.  $J_c$ evolves in time as a result of competing
influences from gravitational damping (which feeds the r-mode), viscosity
(which damps the r-mode) and the magnetic damping term $F_m$.  Both
bulk and shear viscosities are included in our analysis although for
simplicity we keep the star at a uniform temperature of $10^9$K for the
duration of the evolution.  Although viscosities (especially bulk viscosity)
is strongly temperature-dependent, our approximation is not as drastic as it
may seem.  A typical cooling profile of a neutron star, driven by neutrino
emission from the modified URCA process is given by \citep{ Owen98}

\begin{equation}
T_9(t)=\left(\frac{t}{\tau_c}+T_{i,9}^{-6}\right)^{-1/6}\,,
\end{equation}

where $T_9$ is the star's core temperature $T$ in units of $10^9$K, $T_{i,9}$ the initial temperature,  and $\tau_c\approx 1$ yr is the characteristic cooling time from the modified URCA process. If we begin with birth temperatures $T\sim 10^{11}$K, we see that within a few seconds, we have $T\sim 10^9$K. Since the r-mode does not have a large impact until $t\geq 10^2$ seconds, we can approximate $T_9=T/(10^9\,{\rm K})\sim 1$. Subsequent cooling
is on the timescale of years and is also a small perturbation on our results. 
Using the evolution equation for $J_c$, \citep[see eqn.(4)
of][]{Drago1},  one can write

\begin{eqnarray}
\label{spinevo}
\frac{\dot{\Omega}}{\Omega}&=&-2\alpha^2K_c\left[K_jF_g+(1-K_j)[F_v+F_m]\right]-F_{\rm mag} \quad {\rm and}\\ 
\frac{\dot{\alpha}}{\alpha}&=&\left[F_g-[F_v+F_m]\right]-\frac{\dot{\Omega}}{2\Omega}\,, \nonumber
\end{eqnarray}
  
where $F_v$ is the viscous damping rate \citep{Andy1}.  We solve the two coupled equations above numerically. We use the RNS code \citep{stergioulas} to construct 2-dimensional models of rapidly rotating neutron stars. For a given EoS and for a fixed baryonic mass, the RNS code outputs a sequence of neutron star models (with specific gravitational mass, radius, spin frequency etc.) that have increasing central density and decreasing angular velocity. The fastest model in such a sequence spins at or near Kepler frequency. Note that 
the magnetic field does not appear in the RNS code, and in any case, its effect on structure at the field values considered here is negligible. The magnetic field only determines the time to deconfinement, unless the spindown is completely r-mode dominated. As in \citet{Staff1} we assume $5\rho_{0}$ to be a critical density at which quarks in the interior of the neutron star become deconfined - this is not a well determined number, and may span a range from (4-8)$\rho_0$ if a mixed phase of quark and nuclear matter is favored \citep{Glen}. To counter this uncertainty, we have checked our numerical results for a higher putative deconfinement density ($\rho\sim 8\rho_0$) and found that it does not change any of our quantitative conclusions by a significant amount. 

Using the RNS code, we construct sequences of stars (for a given EoS)
with constant baryonic mass and decreasing spin
such that the non-rotating configuration has a central density equal to
$5\rho_{0}$.  We then calculate the time the rotating star takes until its
central density is within $99\%$ of $5\rho_{0}$.  This is a practical way of
using eqn.(\ref{spinevo}) to obtain the minimum mass required (for a given
EoS) to support deconfinement, since the time taken to reach zero
spin is infinite from the magnetic braking being proportional to $\Omega$. 
Since we are interested in the time it takes for a star to spindown to the
critical density for deconfinement, by choosing a sequence where the
non-rotating model has a central density equal to the critical density, this
time gives us the maximum time the star takes to reach the deconfinement
density.

We assume a very small initial amplitude of the r-mode for the fastest spinning star in a sequence, with typical initial $\alpha\sim 10^{-6}$, although our results for those cases where the r-mode saturates are insensitive to this initial value. For the case when the mode does not saturate, we do find mild sensitivity to the initial value of $\alpha$. From the first of eqn.(\ref{spinevo}), we can calculate the time-step between two consecutive neutron star models in the spin-sequence output by RNS code 
\begin{equation}
\label{deltat}
\Delta t \equiv t_{i+1}-t_{i}=\frac{\Omega(t_{i+1})-\Omega(t_i)}{-2K_c\Omega(t_i)\alpha(t_i)^2
[K_j F_g + (1-K_j)(F_v+F_m)] - \Omega(t_i) F_{\rm mag}} \,\,.
\end{equation}

\begin{figure}[h!]
\includegraphics[width=0.5\textwidth]{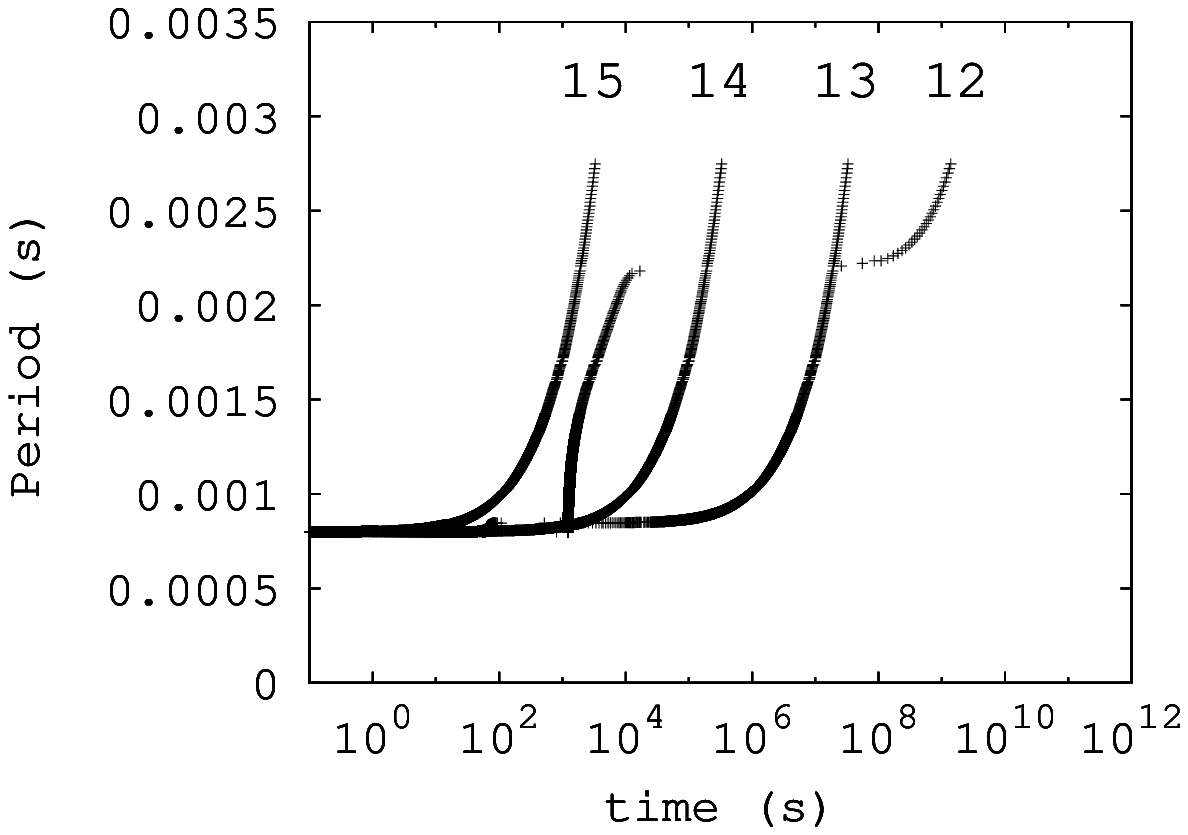}
\includegraphics[width=0.5\textwidth]{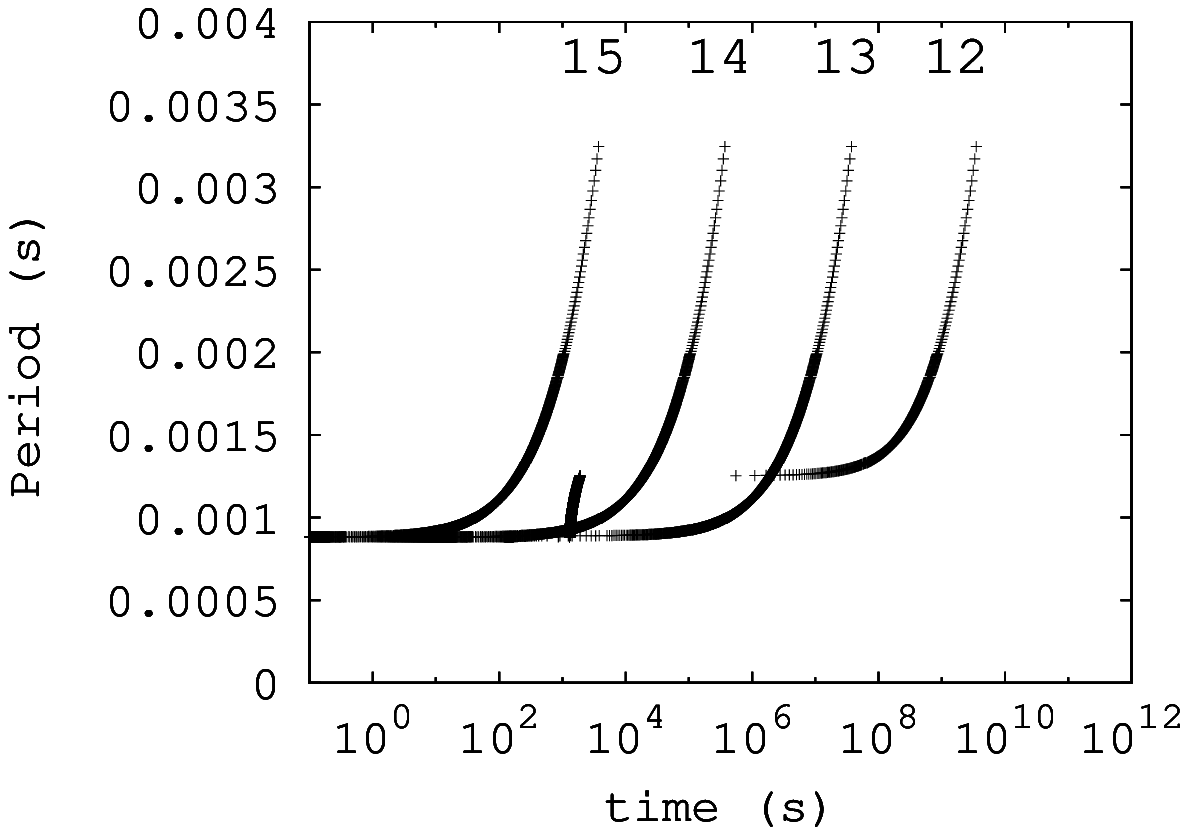}
\includegraphics[width=0.5\textwidth]{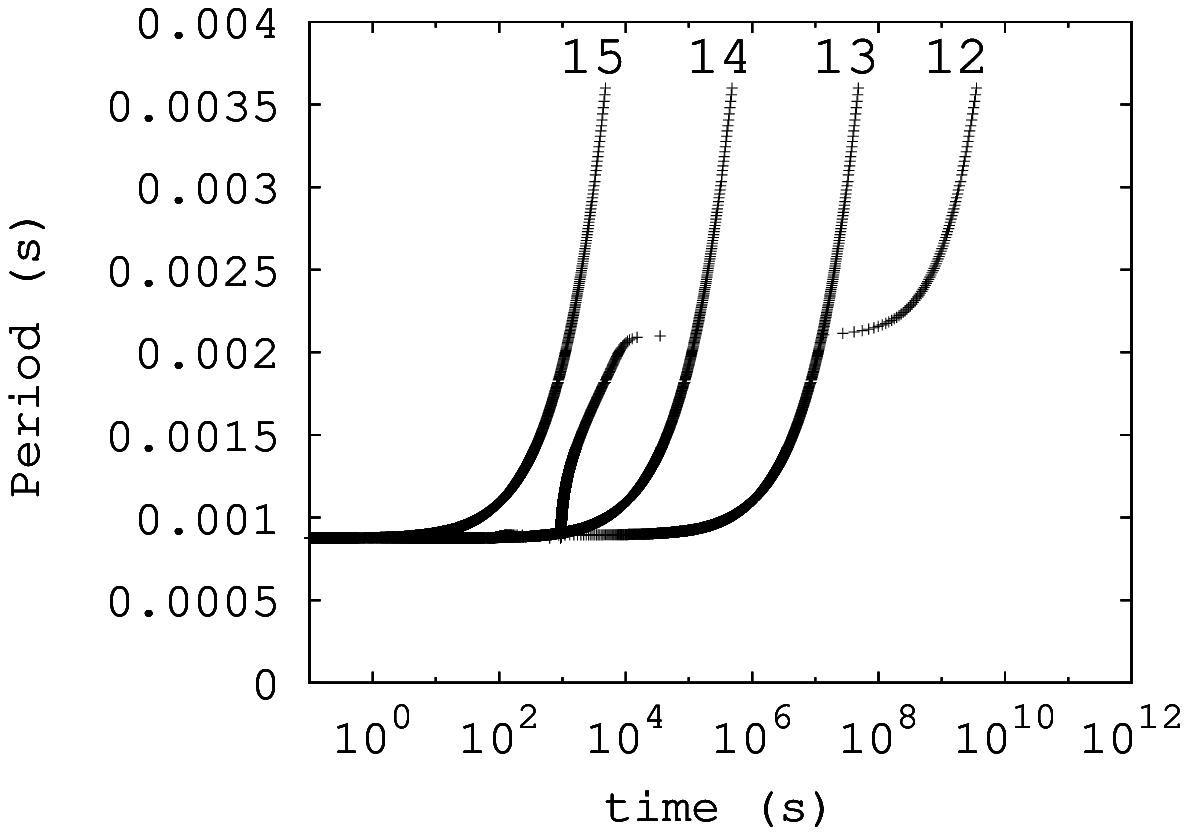}
\caption{Period vs time for the APR EoS ({\it top left}), EoS A ({\it top right}), and EoS BBB2 ({\it bottom left}) with a constant temperature of
$T=10^9$ K. The curves from left to right are labelled for log($B$). For $B\geq 10^{13} G$, magnetic braking starts to dominate the r-mode as far as spindown is concerned. The curves terminate abruptly at the moment when the central density is within $1\%$ of the critical density (see  discussion above eqn.(\ref{deltat})).}
\label{Pvst}
\end{figure}

The inverse dependence on $\alpha$ implies that for small magnetic fields,
where $F_{\rm mag}$ is small, $\Delta t$ is large when the mode amplitude is
just starting to grow.  The RNS code then needs to generate two widely
separated rotating configurations within a sequence, leading to low
resolution in some parts of the $P$ vs $t$ curves for small magnetic fields
(eg., the curve for $B$=$10^{12}$ G in Fig.\ref{Pvst}).  However, this is
only a problem initially and does not affect the total time to
deconfinement.  The above equation, along with the second of eqn.(\ref{spinevo}) is then used to determine the time evolution of $\alpha$. For our
analysis we have used four different magnetic field strengths equally spaced
on a logarithmic scale from $10^{12}$ G to $10^{15}$ G.  We
display results for three different equations of state: EoS A which is composed of only neutrons\citep{Pandharipande} and uses a variational principle to determine the minimum energy state; EoS BBB2 which includes muons and uses a field-theoretic many-body approach \citep{BBB};
and finally EoS APR which is relatively stiff and admits a mixed phase of
quarks and nuclear matter for the heaviest stars \citep{Akmal}.  All three
EoS generate stable configurations for central densities exceeding 5$\rho_0$
and can go up to even $8\rho_0$.  They differ in the details of the density
profile, with the softest (EoS A) having higher interior densities for the
same mass (more compact).  The maximum gravitational mass exceeds
$2M_{\odot}$ only for the APR EoS \citep{Akmal}, with BBB2 providing a
maximum mass of 1.92$M_{\odot}$ \citep{Haensel}.  EoS A is too soft to
generate a 2$M_{\odot}$ neutron star and may seem inadequate to explain recent
observations of such massive compact stars \citep{Demorest}. However, if such massive stars are really quark or hybrid stars, we cannot rule out EoS A in this way, and its inclusion is still useful simply to examine the trend of a relatively soft EoS on the gravitational signal from r-mode driven spindown.

\section{Results: r-mode spindown versus magnetic braking}
\label{Magresults} The following results are obtained on solving eqns.(4)
numerically.  In Fig.~\ref{Pvst}, the curves show how the period,
starting from the Kepler rotation rate at birth, evolves as a function of
magnetic field strength, for different equations of state.  Note that these
curves correspond to structural parameters of that {\it particular} stellar
configuration which {\it just} reaches the putative quark deconfinement
density at zero angular velocity - we may call this the {\it critical
configuration}.  For any given EoS, this is how we determine the minimum
mass and spin period required for a quark phase to appear as a result of
spindown.  Unlike Fig.  4 of \citet{Staff1}, spindown is no longer always
magnetic-dominated for arbitrary magnetic fields.  For example, for
$B\lesssim 10^{12}$ G, r-mode spindown takes over at a few hundred seconds. 
As the instability develops, magnetic damping $F_m$ grows and limits the
growth of the r-mode.  This leads to a plateau in the period that can last
up to several years before magnetic damping once again becomes the dominant
driver of spindown.  It is noteworthy that for the critical configuration,
there is no change in the time to deconfinement for any value of $B$. 
However, if we choose a heavier mass at birth, the time to deconfinement
would be shorter, and can occur while the star is still being spun down as a
result of the r-mode.  Assuming that even the most
massive neutron stars can be transformed to stable quark stars if they reach
the deconfinement density (that is, the quark matter EoS should be
sufficiently stiff to support such a mass), we may expect that several such
neutron stars would have already spun down to the point where they underwent
a quark-hadron phase transition, and now contain quark matter in their core.

Our findings here are different from the conclusions of \citet{Lai} who
studied the r-mode of magnetized and slowly rotating neutron stars, and
found that the r-mode alters the spindown if magnetic fields are less than
$B\sim 10^{14}$G.  Due to the inclusion of the magnetic damping term, which
was omitted in their work, we find that r-mode driven effects on spindown
are pronounced only for $B\lesssim 10^{12}$G, since for higher $B$ fields,
magnetic damping effectively kills the r-mode.  We emphasize that the
effects of magnetic damping for smaller $B$ fields are not as severe as
might be expected from the work of \citet{Shapiro} due to the
self-consistent evolution of $\alpha$ in the presence of the magnetic
damping term, as explained at the beginning of section \ref{Theorymag}.

We also find that the r-mode driven spindown is more pronounced in a
relatively stiff equation of state such as EoS APR.  This is because $F_g$,
the gravitational damping rate is larger for a stiffer EoS~\footnote{
$F_g\propto MR^6/P^6$ and for stars with the same baryonic mass and central
density at zero angular frequency, this quantity is systematically larger
for a stiffer EoS.  In other words, for a given baryonic mass, a stiffer EoS
can support a larger gravitational mass and radius, and has a slightly
smaller Kepler frequency since its average density is smaller (less
compact).  For eg., $F_g^{\rm APR}/F_g^{\rm EoS A}\approx$ 6.}, driving the
r-mode unstable in a shorter amount of time (effectively spinning the star
down to deconfinement density quickly).

The saturation amplitude $\alpha_{\rm sat}$ of the r-mode assumed in the
foregoing analysis is ${\cal O}(1)$.  Such a large value is probably
unrealistic due to multi-mode coupling and onset of non-linear effects
\citep{Arras02,Bondarescu08}.  If we use a smaller saturation value, we find
that our conclusions change quantitatively.  The r-mode grows more slowly
for a smaller $\alpha_{\rm sat}$, but continues to be important since the
main damping agent $F_m$, which is roughly proportional to $\alpha^4$, is
also smaller.  As a consequence, the spindown due to the r-mode is weaker,
but also lasts longer, as seen in Fig.\ref{Pvst0.01}.  This can lead to the
star taking a shorter time to reach a given period in the case of smaller
$B$ fields.  This can be seen clearly in Fig.\ref{Pvslogalpha005} which
compares the time taken by the star to spindown to a period $P$=3 ms, for
$\alpha_{\rm sat}$=0.01 and $\alpha_{\rm sat}$=0.005.

\section{Evolution of the r-mode and Gravitational Wave Frequency}
\label{GWevo}
As shown in the previous section, fairly large magnetic fields of $10^{13}$G or more are required to make the r-mode irrelevant for spindown. For smaller magnetic fields, the r-mode evolution can be the main agent for spinning down rapidly rotating stars. Following \citet{Owen98}, we can obtain expressions for the growth of the mode amplitude ($\alpha$) and the corresponding evolution of the frequency of the gravitational wave ($f$) under certain approximations.

\subsection{Saturated Phase}
The r-mode amplitude grows rapidly until non-linear saturation occurs \citep{Arras02,Bondarescu07} at which point $\dot{\alpha}$=0. We can obtain a description of this saturated phase from eqns.(\ref{spinevo}), which imply that the rotation rate evolves according to
\begin{equation}
\label{sateqn}
\frac{\dot{\Omega}}{\Omega}=\frac{2\alpha^2K_cF_g-F_{\rm mag}}{1-\alpha^2K_c(1-K_j)} \,,
\end{equation}
where $\alpha=\alpha_{\rm sat}$ is the saturation amplitude of the r-mode. Note that $F_m$ does not appear since it can be eliminated using the second line of eqns.(4). Its effect will however show up in the growth phase discussed below. Using the fact that $\Omega=3\pi f/2$ for the $l=m=2$ r-mode, setting $K_c=0.1$ (a typical value determined from eqn.(\ref{IandJbar})),  and using the frequency-dependent expression for the gravitational wave damping timescale $\tau_{\rm GR}\equiv F_g^{-1}$, we find

\begin{equation}
\label{ftsat}
\frac{\dot{f}}{{\rm Hz}^2}\approx -0.35\alpha^2\left(\frac{f}{1 {\rm kHz}}\right)^7-0.0055\frac{B_{14}^2R_6^6}{I_{45}}\left(\frac{f}{1 {\rm kHz}}\right)^{3} \,,
\end{equation}

 where we have used the expression for $\tau_{\rm GR}$ from
\citet{Morsink} and neglected $\alpha^2K_c(1-K_j)\ll 1$ in the denominator
of eqn.(\ref{sateqn}).  $B, R$ and $I$ are expressed in terms of reduced
dimensionless units.

\vskip 0.5cm
\subsection{Growth Phase}
In the initial stages of the mode growth phase, the viscosity controls the evolution of the r-mode, but the instability to gravitational waves soon takes
over. Viscous damping from bulk viscosity can be large at $T\geq 10^{11}$K,
but as the neutron star cools rapidly on the order of seconds, our approximation of setting $T_9=1$ implies that viscosity does not affect the subsequent spindown behaviour or the gravitational waveform during the growth of the r-mode instability. It follows from eqns.(\ref{spinevo}) that during this phase,
since $\alpha$ is small

\begin{equation}
\frac{\dot{\alpha}}{\alpha}=(F_g-F_m)-\frac{\dot{\Omega}}{2\Omega}\,\,.
\end{equation}

The angular velocity evolves according to the first line of eqns.(4).
We assume $K_j\approx 0$ for the growth phase, which is tantamount to including fully the canonical angular momentum of the r-mode in the star's physical angular momentum. This is only justified if differential rotation is small \citep{Sa}, as is the case when the mode is still small but growing. Then,
as before, the proportionality between $f$ and $\Omega$ implies 

\begin{equation}
\label{ftgrow}
\frac{\dot{f}}{{\rm Hz}^2}\approx - 0.0131\alpha^2(t)\left(\frac{f}{1 {\rm kHz}}\right)^3-0.0055\frac{B_{14}^2R_6^6}{I_{45}}\left(\frac{f}{1 {\rm kHz}}\right)^{3}-\frac{0.202B_{14}^2R_6}{\bar{J}M_{1.4}}\alpha^2(t)\int_0^t\alpha^2(t^{\prime})\left(\frac{f(t^{\prime})}{{\rm Hz}}\right)dt^{\prime}\,.
\end{equation}

 where we have used the expression for $F_m$ from eqn.(11) of \citet{Drago1}. We can now use eqns.(\ref{ftsat}) and (\ref{ftgrow}) to obtain the gravitational strain amplitudes and waveforms for the saturated phase and growth phase respectively. 

\section{Results: Gravitational Strain Amplitudes and Waveforms}
\label{GWresults}
\begin{figure}[h!]
\includegraphics[width=0.5\textwidth]{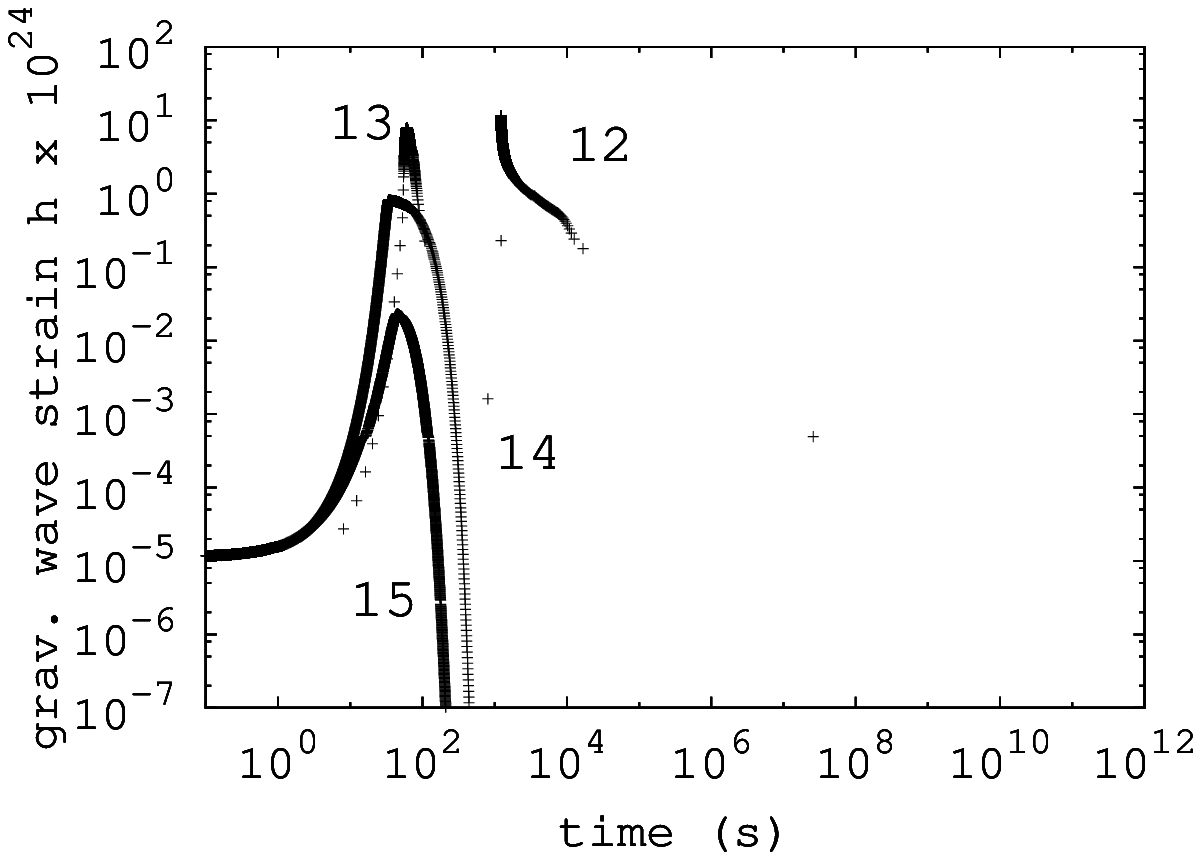}
\includegraphics[width=0.5\textwidth]{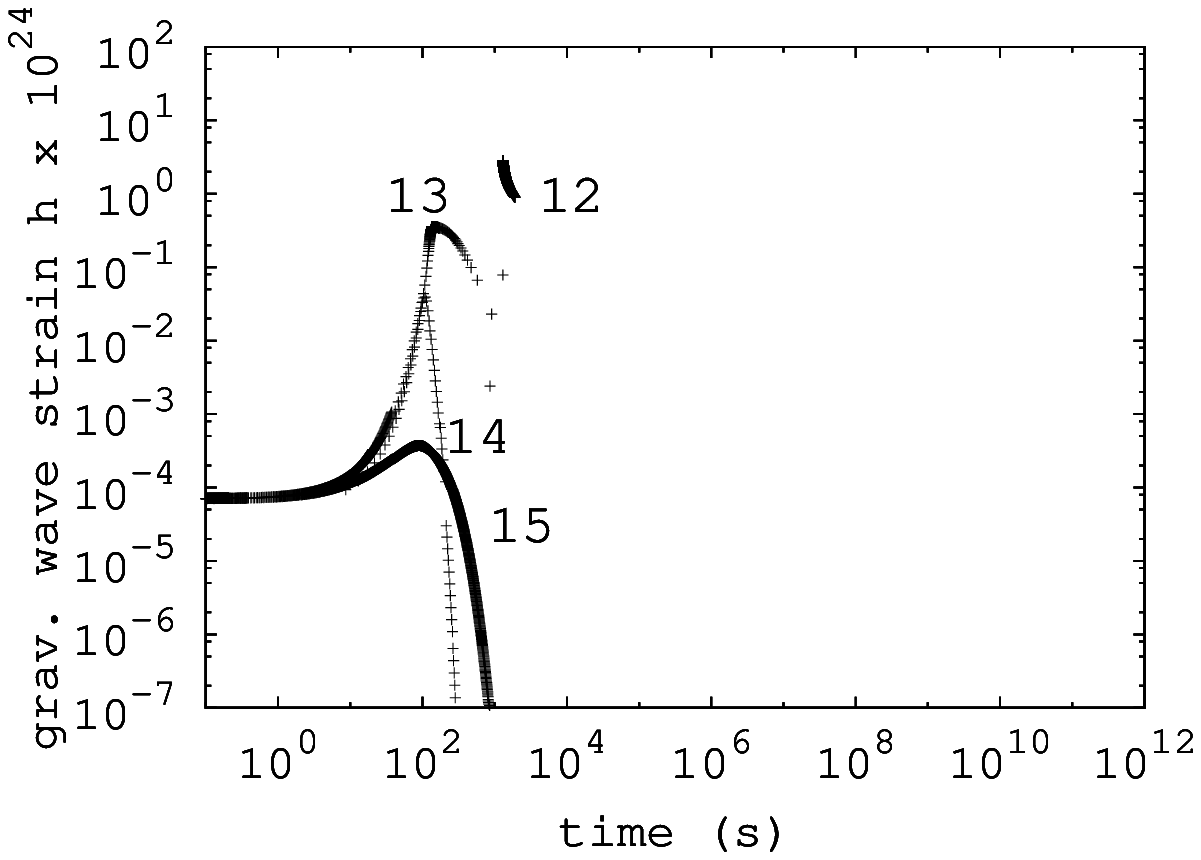}
\includegraphics[width=0.5\textwidth]{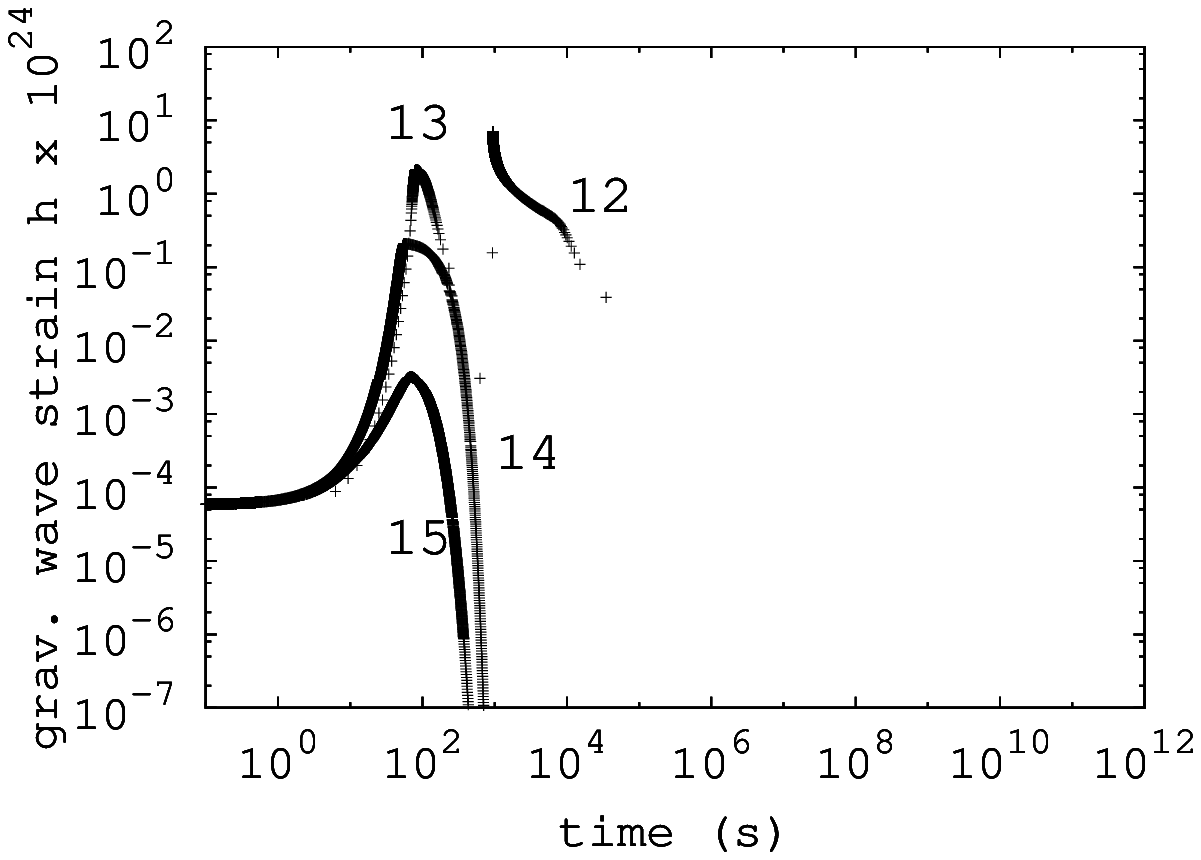}
\caption{Gravitational wave strain as a function of time for  the APR EoS
({\it top left}), EoS A ({\it top right}),
and EoS BBB2 ({\it bottom left}) with a constant temperature of
$T=10^9$K. The curves for softer EoS and larger $B$ fields have a much lower peak strain as in these cases the r-mode amplitude never saturates.}
\label{gwvst}
\end{figure}

The strain amplitude $h(t)$ 
corresponding to the $l$=$m$=2 mode is found by a standard multipole
analysis \citep{Thorne}. Including the angle-average over the 
position of sources in the sky,

\begin{eqnarray}
\label{ht}
h(t)&=&\sqrt{\frac{3}{80\pi}}\frac{\omega^2S_{22}}{D} \,,\\
S_{22}&=&\sqrt{2}\frac{32\pi}{15}\frac{GM}{c^5}\alpha\Omega R^3\bar{J} \,.
\end{eqnarray}

where $S_{22}$ is the current multipole as given by eqn.(3.9) of
\citet{Owen98}, the mode frequency $\omega=4\Omega/3$ and $D$ is the source
distance, chosen henceforth to have a typical value of 20 Mpc~\footnote{This
fiducial value of $D$ extending to the Virgo cluster is chosen to encompass
enough neutron stars to ensure a reasonable event rate \citep{Owen98}.}. 
Figure \ref{gwvst} shows the time-evolution of the strain amplitude $h(t)$
for various magnetic fields and for different EoS for $\alpha_{\rm sat}=0.5$,
while Fig.~\ref{gwvst0.01} shows this for $\alpha_{\rm sat}=0.01$.  For magnetic fields
$B\geq 10^{14}$G, the r-mode withers much before it nears saturation, for
any EoS - the strain amplitude is consequently very small.  For $B\sim
10^{13}-10^{14}$G, we notice a strong dependence on the equation of state. 
The soft EoS A does not lead to a saturating r-mode while relatively stiffer
EoS do.  For weaker magnetic fields around $B\sim 10^{12}$G, the r-mode
saturates and displays the behaviour shown in Fig.  5 of \citet{Owen98}.

We now examine the gravitational waveform in the frequency domain $\tilde{h}(f)$ which is the Fourier transform of the time signal. 
This is useful in estimating the signal-to-noise ratio (SNR) in matched filtering techniques. In the stationary phase approximation,

\begin{equation}
|\tilde{h}(f)|=\sqrt{\frac{|h(t)|^2}{|\dot{f}|}} \,\,.
\label{htildeqn}
\end{equation}

\begin{figure}[h!]
\includegraphics[width=0.5\textwidth]{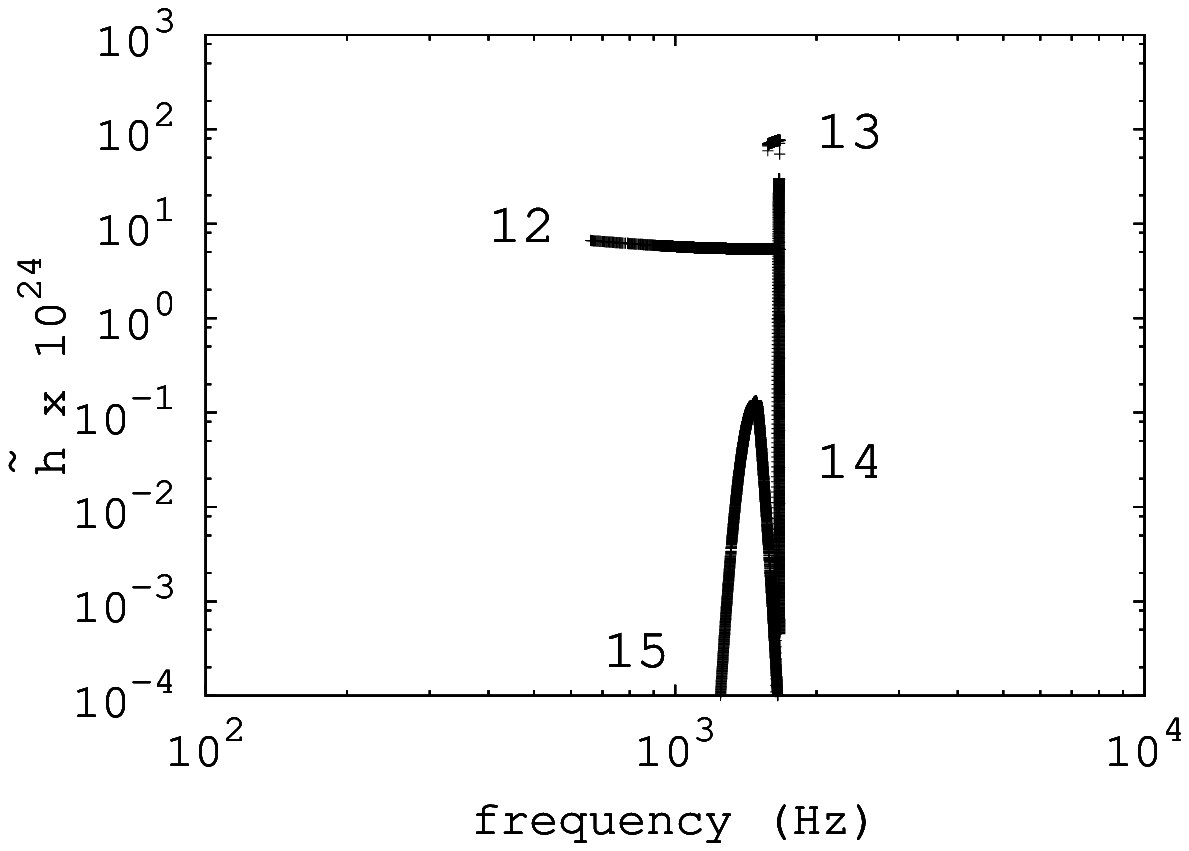}
\includegraphics[width=0.5\textwidth]{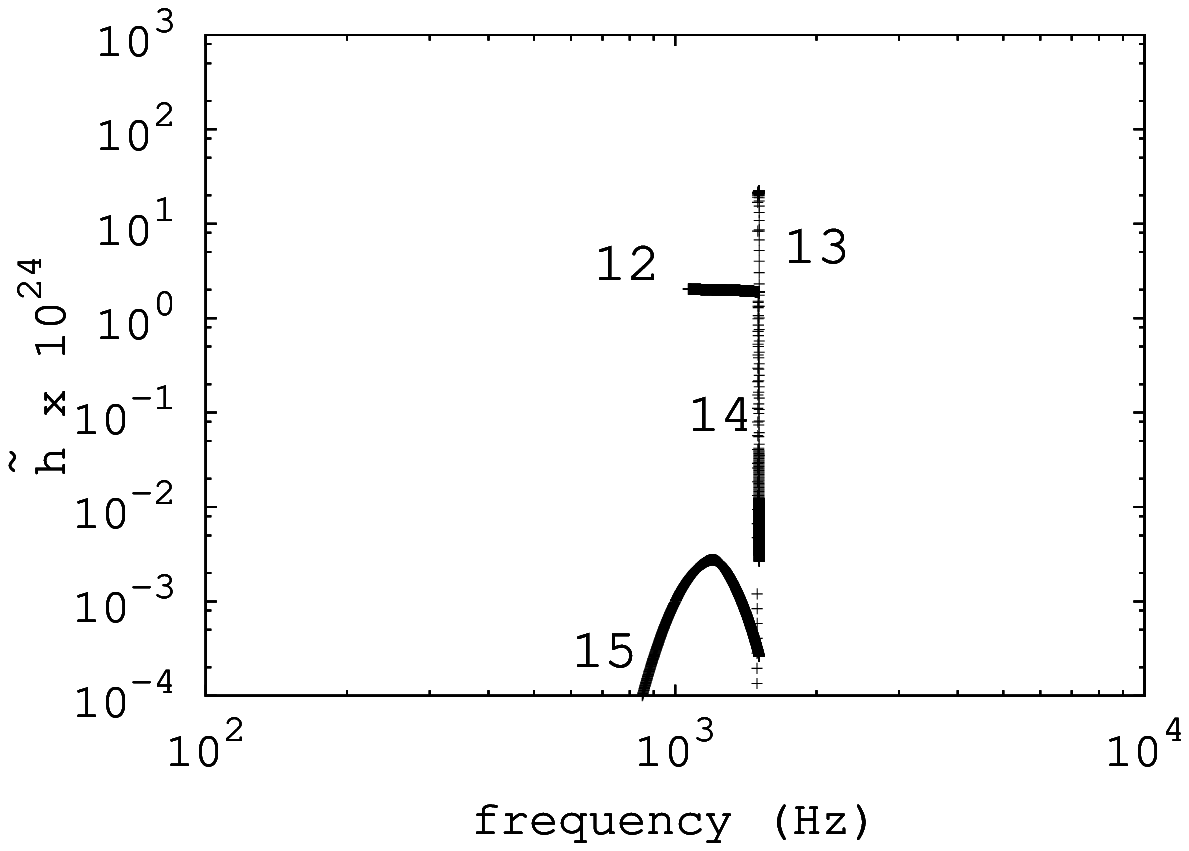}
\includegraphics[width=0.5\textwidth]{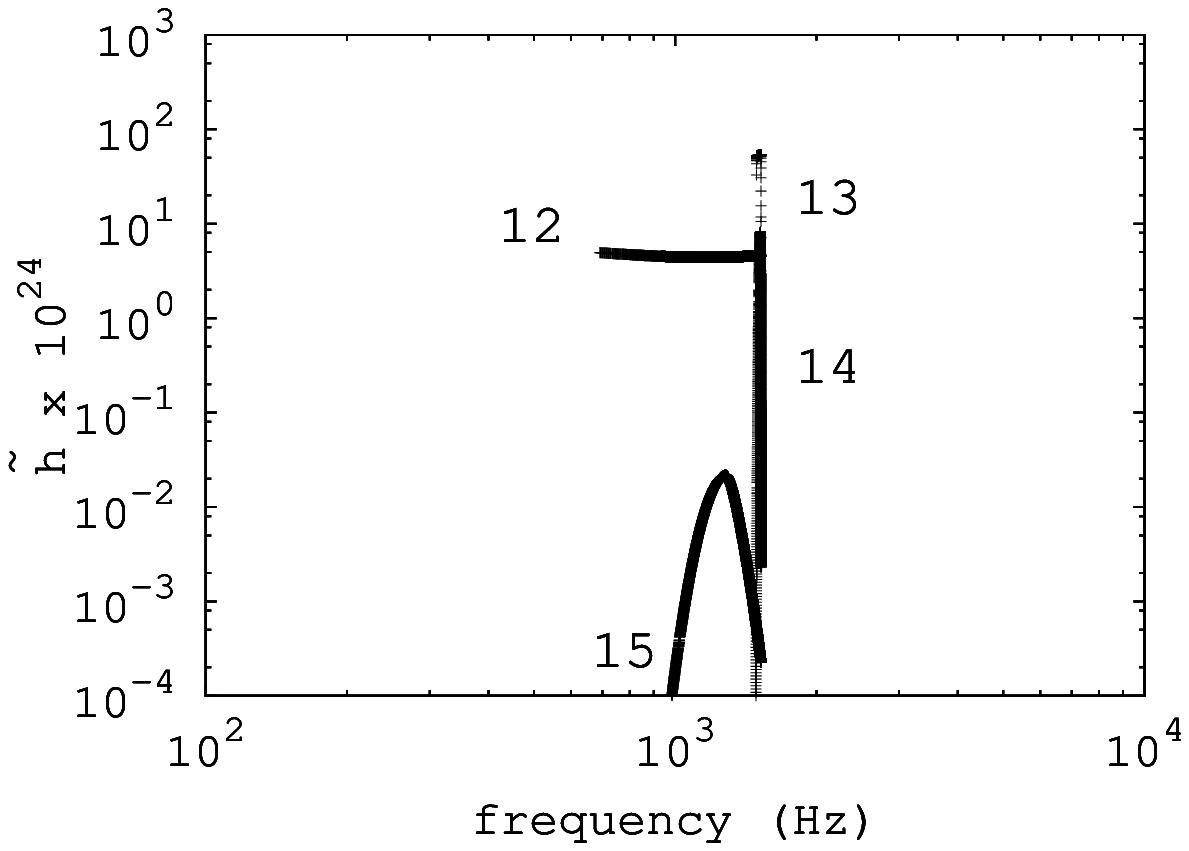}
\caption{Gravitational wave strain as a function of frequency for  the APR EoS
({\it top left}), EoS A ({\it top right}),
and EoS BBB2 ({\it bottom left}) with a constant temperature of
$T=10^9$K. The $10^{14}$ G and $10^{15}$ G curves are distinct from the rest, as in these cases the r-mode amplitude barely or never saturates.}
\label{htilde}
\end{figure}

From eqns.(\ref{ftsat}), (\ref{ftgrow}) and (\ref{ht}), we can find $\tilde{h}(f)$ for both the growth phase and the saturated phase. 
 From Figs.~\ref{htilde} and \ref{htilde0.01} (for $\alpha_{\rm
sat}=0.5$ and $0.01$ respectively), we see that for magnetic fields $B\leq 10^{13}$G,
the signal from the r-mode is similar to previous result obtained in the
absence of magnetic fields \citep{Owen98}. A sharp peak at high frequency ($\approx$ 1.6 kHz) corresponding to the growth phase is seen, followed by a plateau signal at lower frequencies as the r-mode saturates. Note that the curves terminate on the left since at that point, deconfinement density is reached and the subsequent signal from the phase transition needs a detailed analysis beyond the scope of this work. However, we refer the reader to the Appendix for a preliminary treatment of this issue. Previous works \citep{Schenk02,Morsink02,Brink,Bondarescu07} indicate that the saturation amplitude $\alpha_{\rm sat}$ may be quite small, around $10^{-2}$ instead of order 1. Still, it is useful to contrast both cases: when $\alpha_{\rm sat}\approx 0.5$ and $\alpha_{\rm sat}\approx 0.01$.

\subsection{Signal Detectability}

\begin{figure}[h!]
\includegraphics[width=0.5\textwidth]{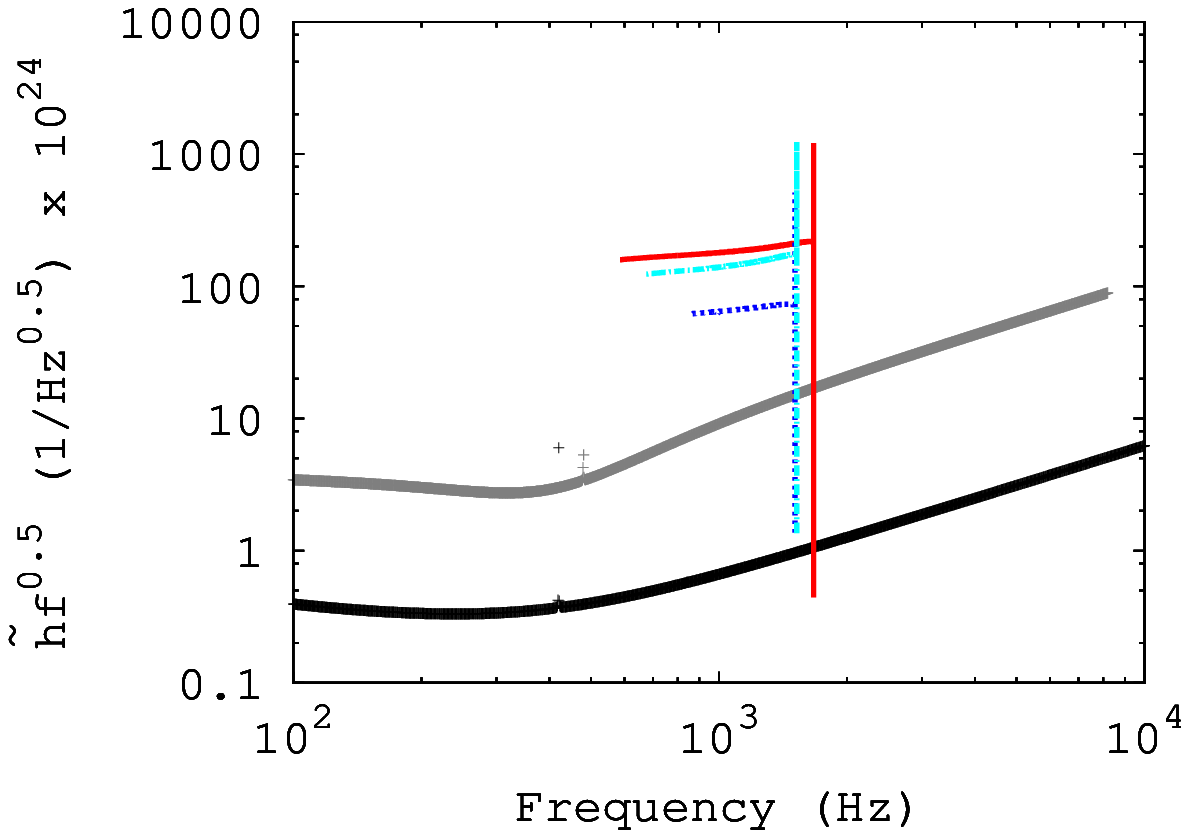}
\includegraphics[width=0.5\textwidth]{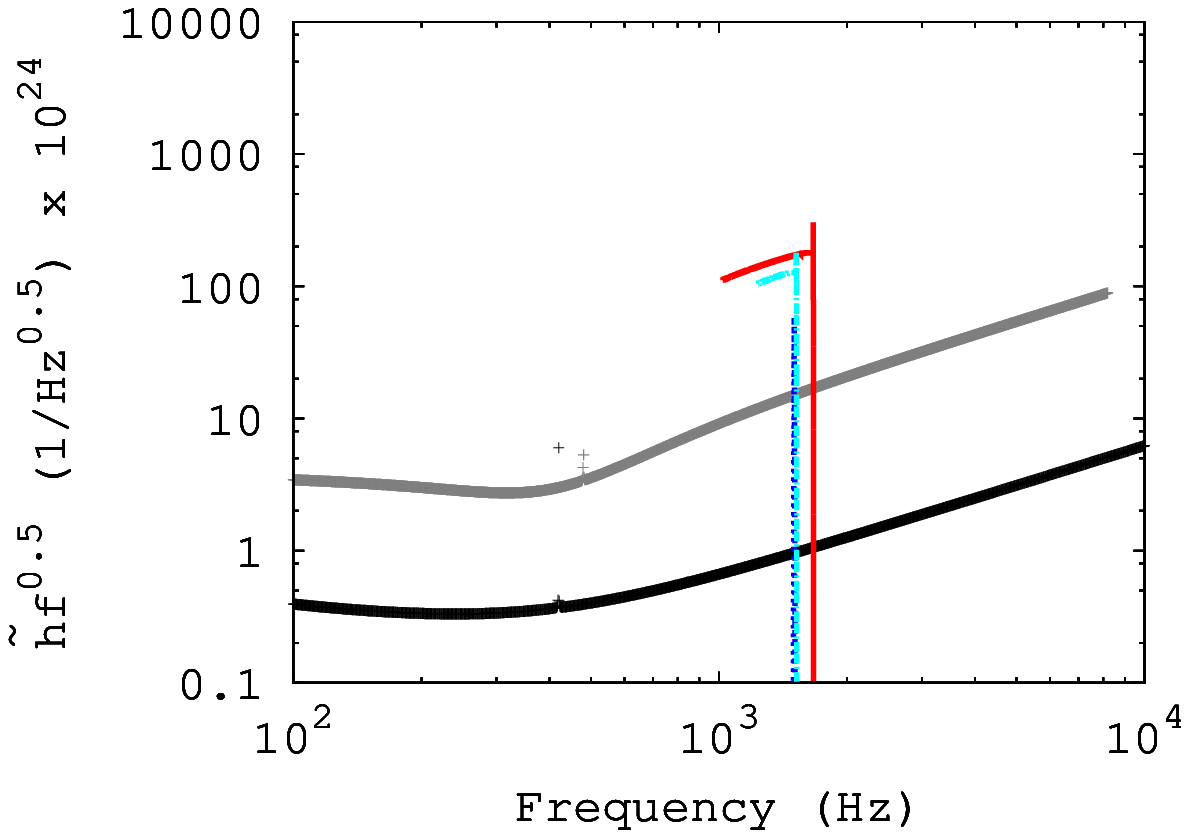}
\caption{Gravitational wave strain as a function of frequency for three different EoS - APR(solid), BBB2(dash-dotted) and EoS
A(dotted), compared against sensitivity curves (strain noise) for (grey) Advanced LIGO taken from \citet{Harry} for the LIGO Scientific Collaboration, and (black) the Einstein Telescope from \citet{hild10}. Left panel is with B=$10^{13}$ G and right panel with B=$10^{14}$ G. The saturation amplitude $\alpha_{\rm sat}$=0.01.}
\label{hf-B}
\end{figure}

\begin{figure}[h!]
\includegraphics[width=0.5\textwidth]{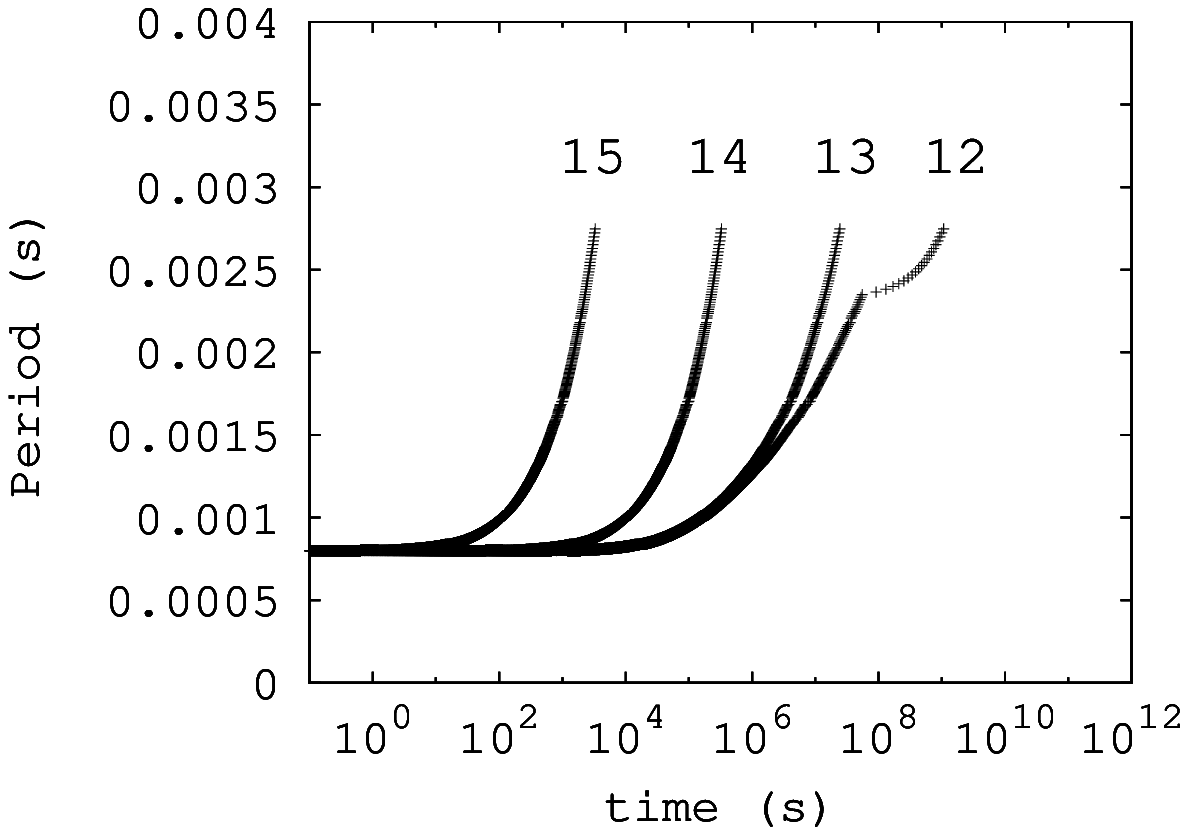}
\includegraphics[width=0.5\textwidth]{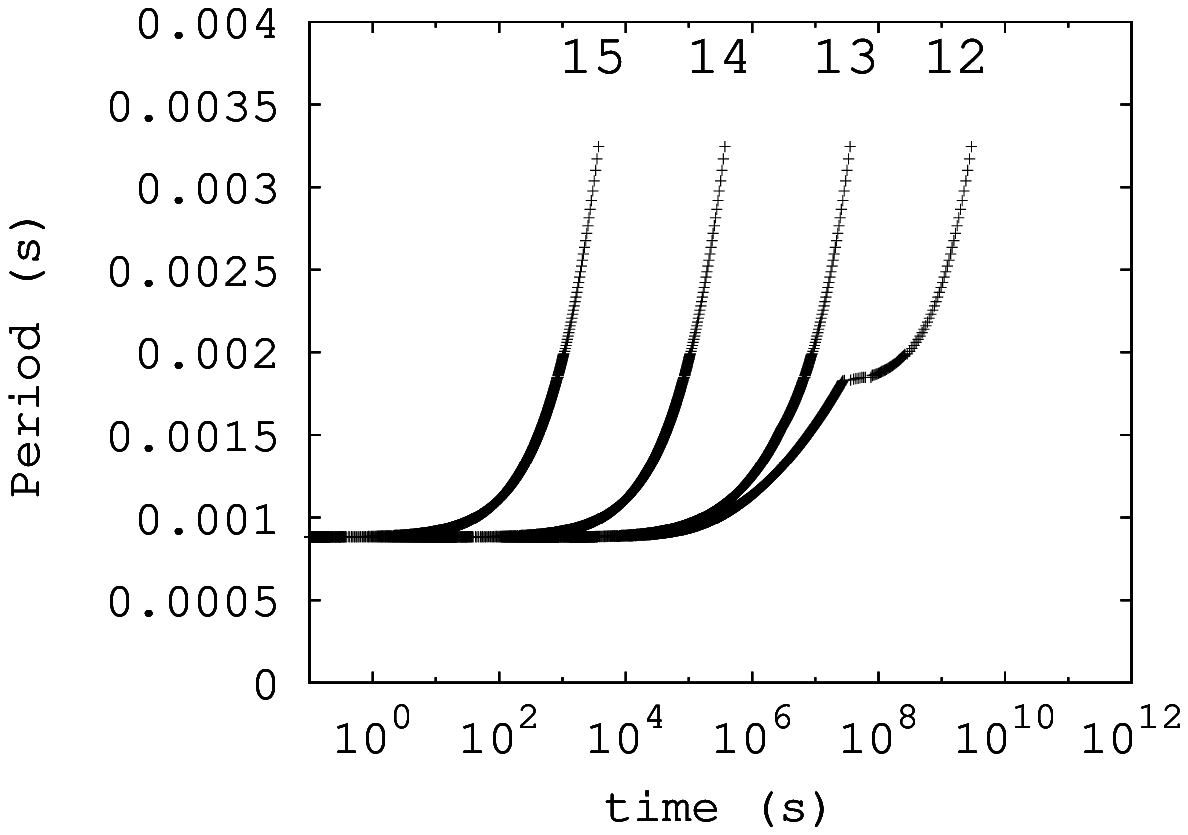}
\includegraphics[width=0.5\textwidth]{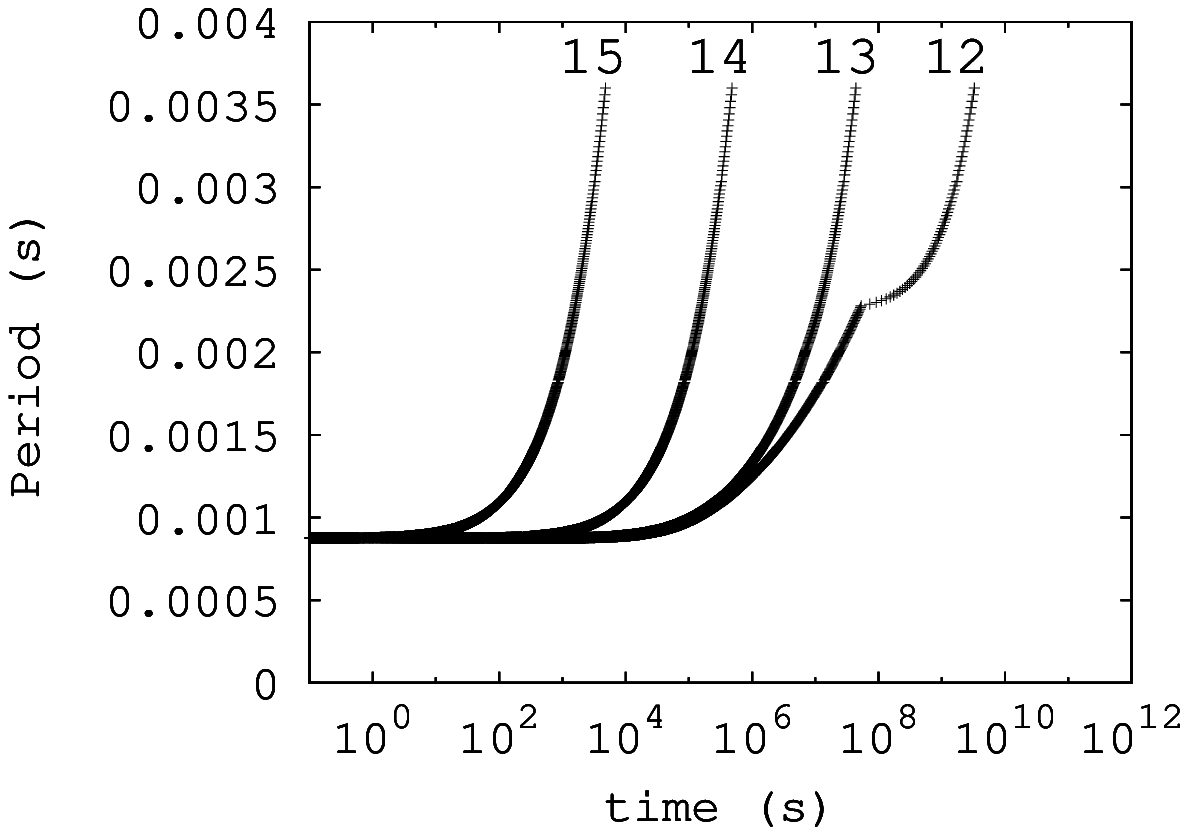}
\caption{Same as Fig.~\ref{Pvst}, but with r-mode saturation value $\alpha_{\rm sat}=0.01$. Note that the effect of the r-mode is still significant but appears later than for $\alpha_{\rm sat}=0.5$. This is due to the nature of magnetic damping.
}
\label{Pvst0.01}
\end{figure}

\begin{figure}[h!]
\includegraphics[width=0.5\textwidth]{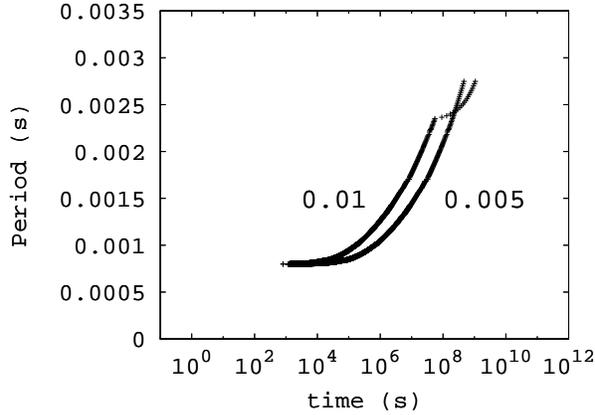}
\caption{Spindown from Kepler frequency for identical stellar configurations (APR EoS with $B$=$10^{12}$ G) but with different r-mode saturation values: $\alpha_{\rm sat}$=0.01 and 0.005. Time to reach $P$=3 ms (where supposed deconfinement density 5$\rho_0$ is reached) is shorter for smaller $\alpha_{\rm sat}$.
}
\label{Pvslogalpha005}
\end{figure}

\begin{figure}[h!]
\includegraphics[width=0.5\textwidth]{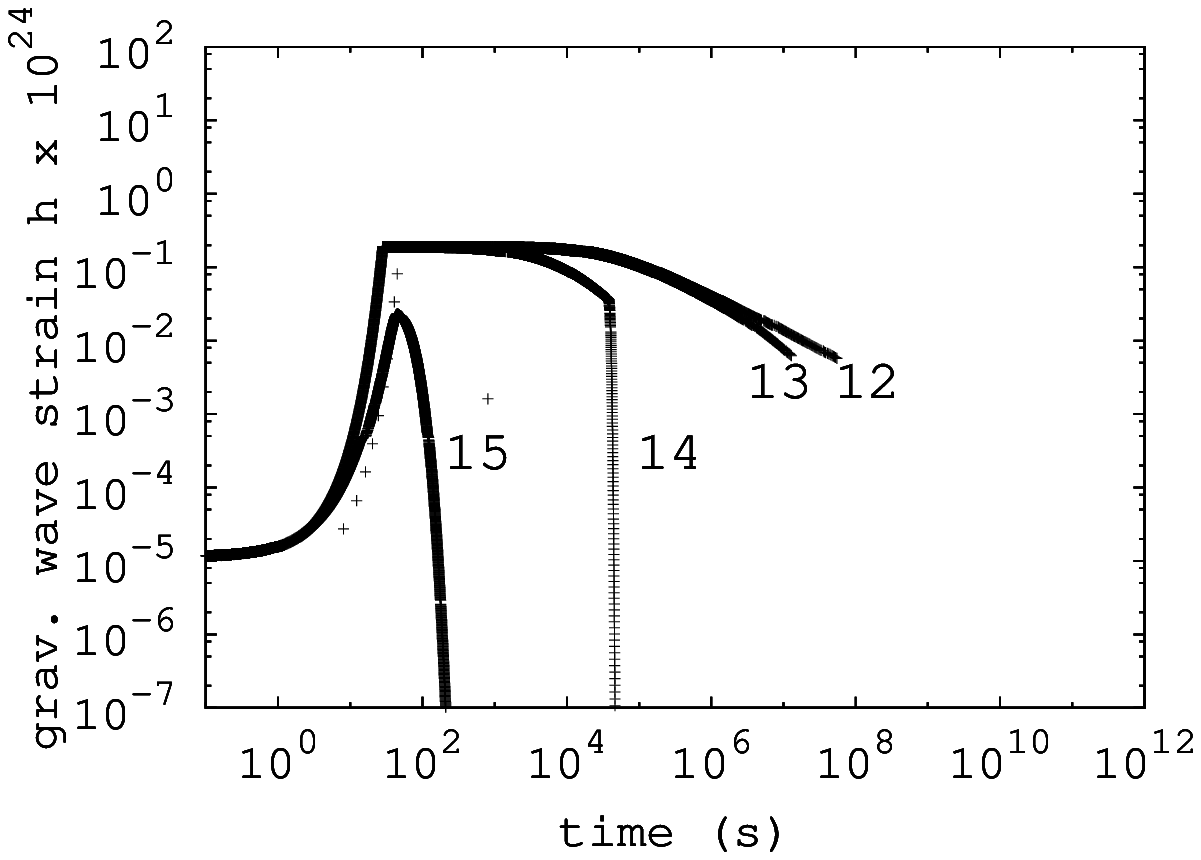}
\includegraphics[width=0.5\textwidth]{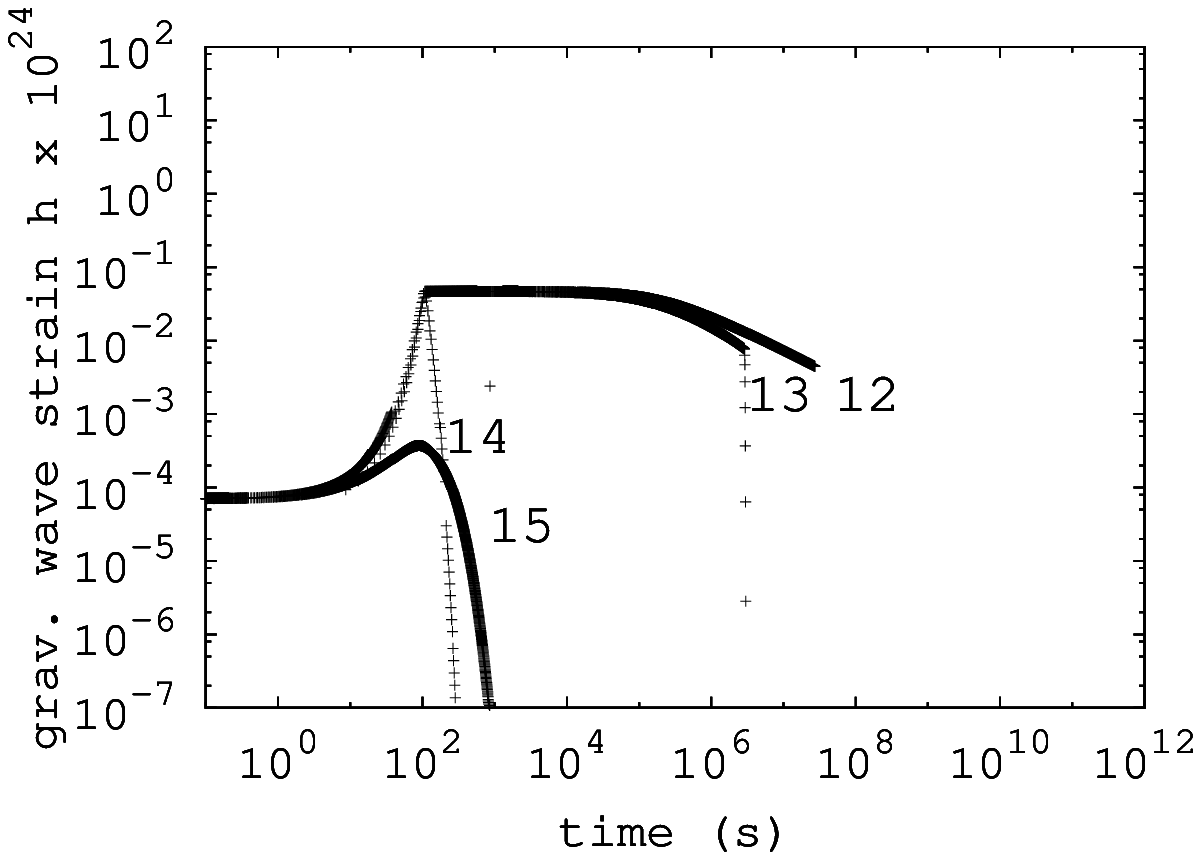}
\includegraphics[width=0.5\textwidth]{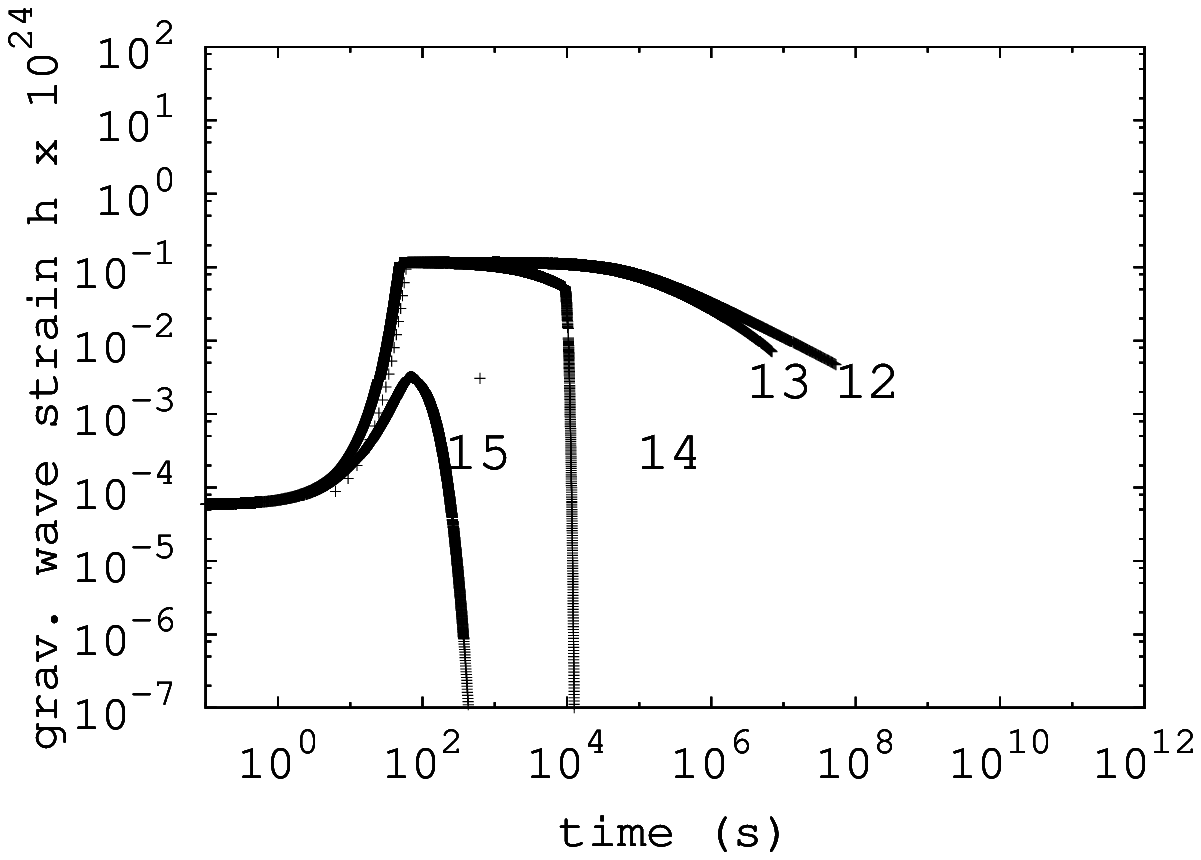}
\caption{Same as Fig.~\ref{gwvst} but with $\alpha_{\rm sat}=0.01$. 
For smaller values of $B$, the signal reflects the fact that the r-mode evolves on much longer timescales than for $\alpha_{\rm sat}=0.5$.}
\label{gwvst0.01}
\end{figure}

\begin{figure}[h!]
\includegraphics[width=0.5\textwidth]{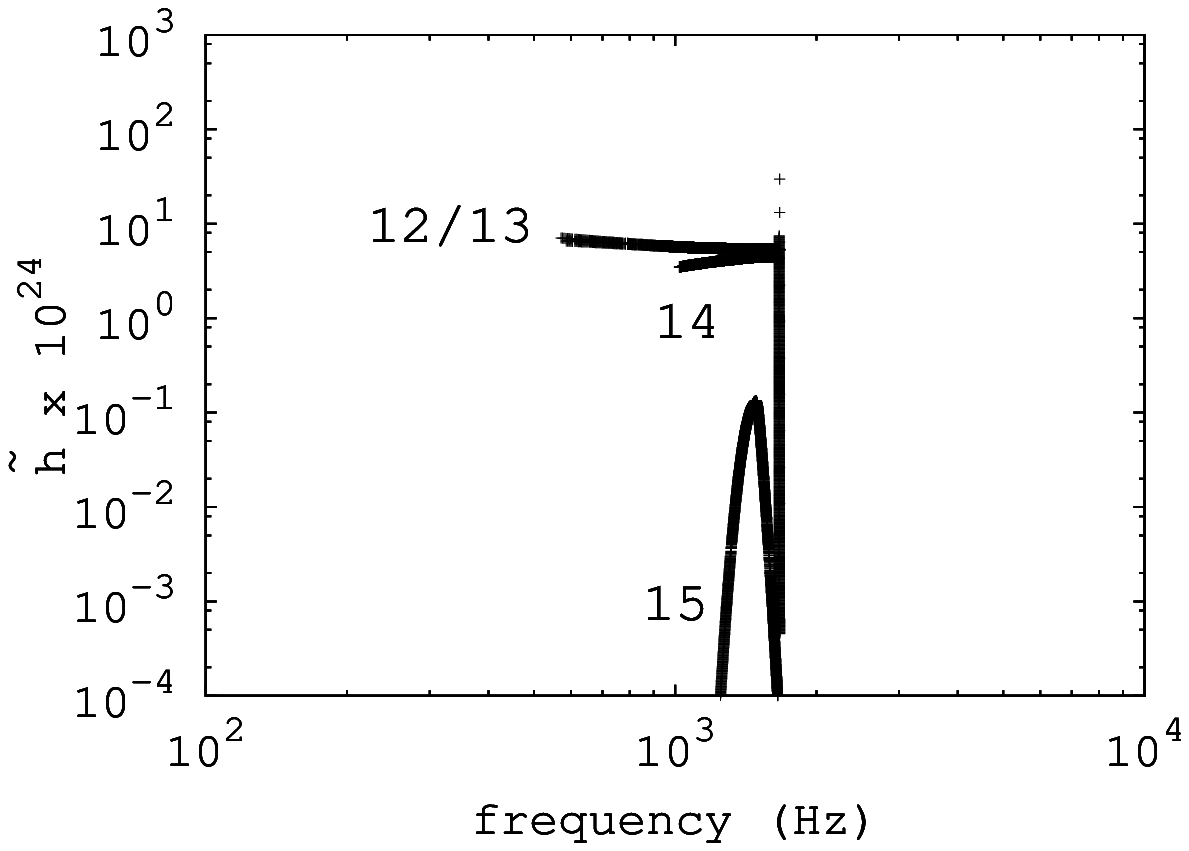}
\includegraphics[width=0.5\textwidth]{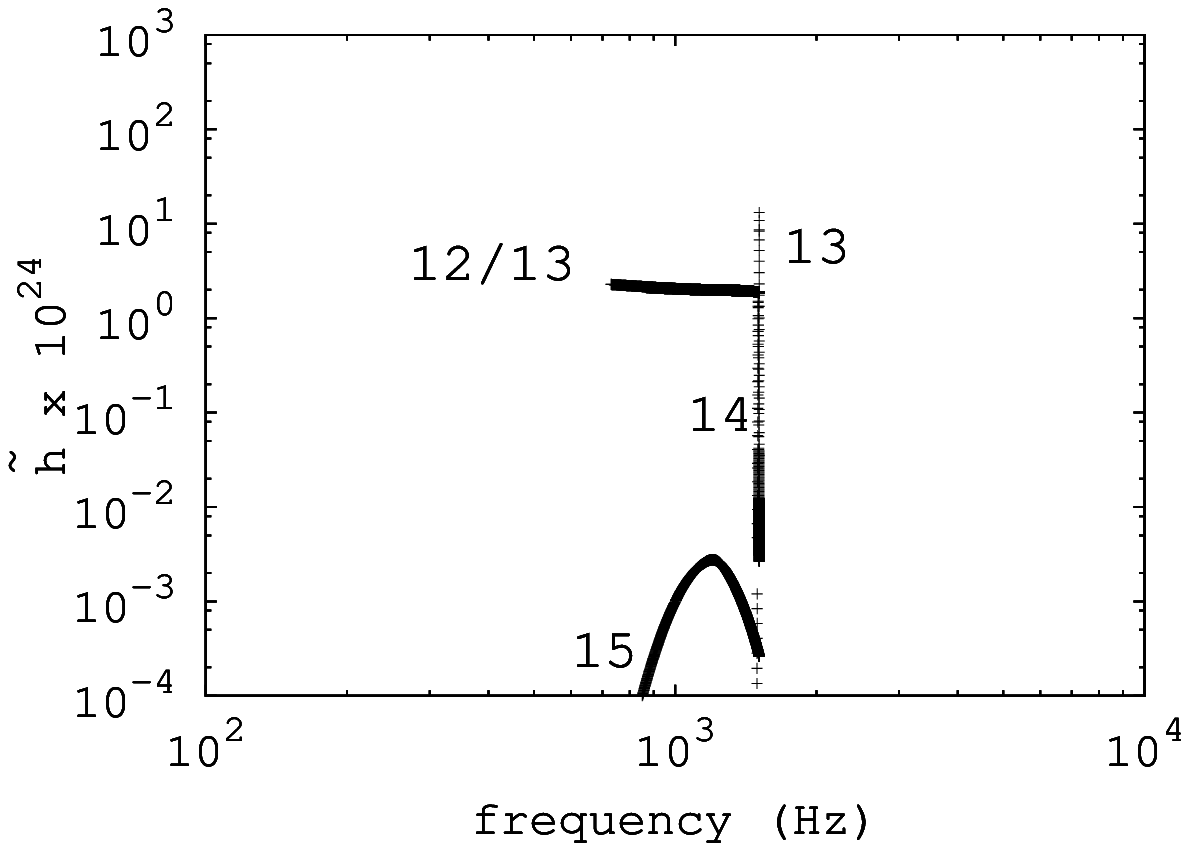}
\includegraphics[width=0.5\textwidth]{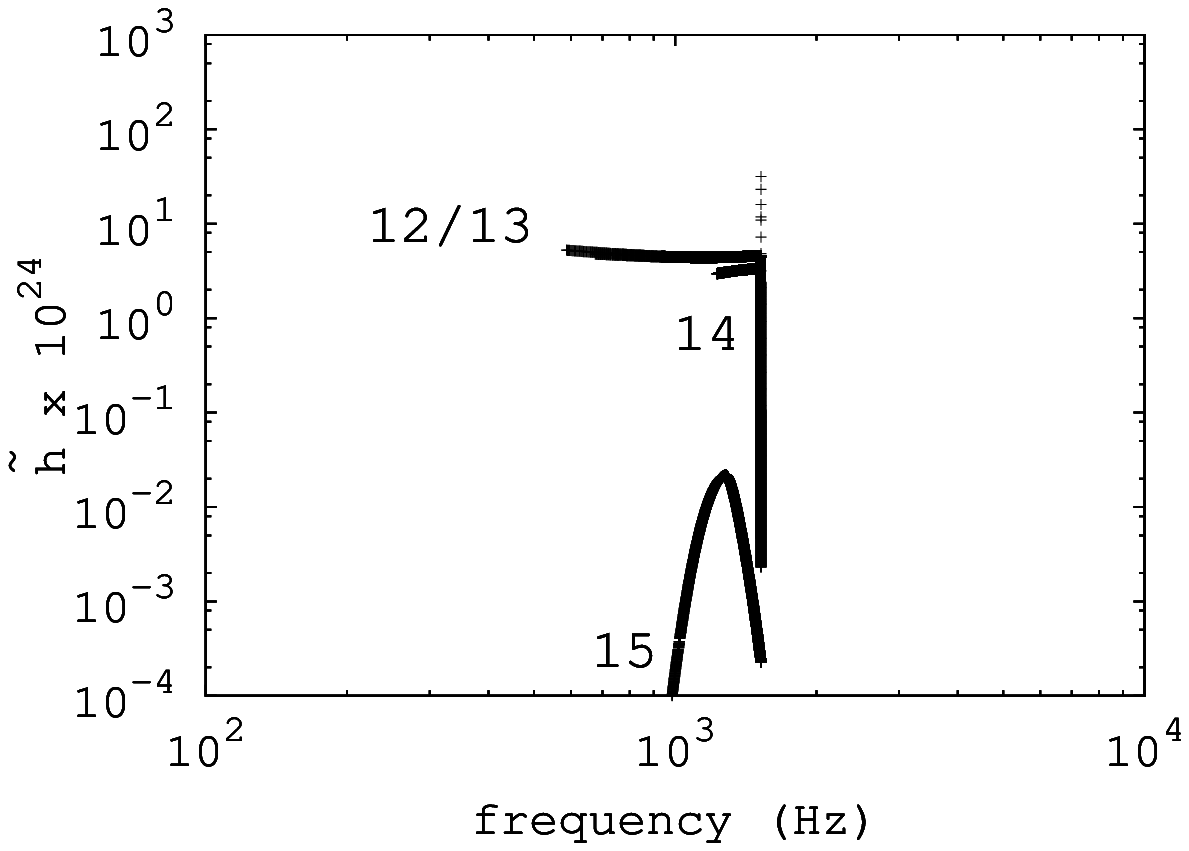}
\caption{Same as Fig.~\ref{htilde}, but with $\alpha_{\rm sat}=0.01$. 
The $10^{15}$ G (and for EoS A also the $10^{14}$ G) are distinguished, as 
in these cases the r-mode amplitude never saturates even when we assume such a low saturation value.}
\label{htilde0.01}
\end{figure}

In Fig.~\ref{hf-B} we present our estimate of the continuous gravitational
wave signal from the spindown (assuming a point source at a distance of 20
Mpc), compared to the anticipated sensitivity of Advanced
LIGO~\footnote{These curves represent the incoherent sum of the principal
noise sources, such as quantum noise, seismic noise and thermal noise, as
best understood at this time.  There will be, in addition, technical noise
sources.  These curves serve as a guide to the overall curve and an early
approximation to the anticipated sensitivity \citep{LIGO}.} and the Einstein
telescope \citep{hild10}.   The left panel of Fig.~\ref{hf-B} shows the
weighted strain amplitude (1/$\sqrt{\rm Hz}$) versus frequency for the three EoS studied in this paper at $10^{13}$ G, compared to the anticipated noise-weighted design sensitivities of Advanced LIGO and the Einstein telescope. The right panel shows the same for $10^{14}$ G.  In both figures, we are assuming
$\alpha_{\rm sat}\approx 0.01$.  The saturated phase of the curve is at
least an order of magnitude above the anticipated sensitivity for Advanced
LIGO for all EoS at $10^{13}$ G and could therefore be expected to be
observed out to 20 Mpc with Advanced LIGO, with a typical SNR of $\sim$ 20. 
At $10^{14}$ G and for the softest EoS, the SNR in Advanced LIGO drops to
$\sim$ 2. The Einstein telescope is expected to gain about an order of
magnitude in sensitivity compared to Advanced LIGO, and hence all of the
above cases should be detectable in principle. Our SNR estimates only provide upper limits since matched filtering is probably impractical for such sources, where spin parameters cannot be promptly measured, though the source location could be known if the supernova is observed.  However, we note that most of the signal comes from the saturated phase, which for the case of small $\alpha_{\rm sat}$ occurs several years after a neutron star's birth in a supernova.  Observations of the neutron star's spindown parameters may then be possible, making the matched filtering method more feasible. In the context of r-modes in newly-born neutron stars, methods other than matched filtering have been suggested~\citep{Brady,Owen,Zhu} that would be less computationally intensive.

\subsection{Saturation at $\alpha_{\rm sat}\approx0.5$}

As mentioned in \citet{Owen98}, the sharp spike in the strain amplitude seen
in certain cases only lasts for a brief period of time (of the order of a
minute) and carries a small fraction of the total energy emitted in
gravitational waves over the spindown epoch (see Fig.~\ref{gwvst}). 
However, for the softer EoS, such as BBB2, we see that for magnetic fields
$B\approx 10^{15}$G, the spike is softened into an asymmetric hump.  In this
case, the growth phase lasts about an hour and carries a larger fraction of
the emitted energy - it may be detectable owing to its larger SNR in second
and third generation detectors.  Furthermore, for the softest EoS considered
here (EoS A), there is no saturation regime for fields larger than $B\approx
10^{12}$ G.  These are potentially distinguishing feature of the equation of
state in gravitational waves.  In general, for very high magnetic fields
$B\geq 10^{15}$ G, the r-mode is strongly suppressed, and the signal in
gravitational waves is weak, with little possibility of detection even in
third generation detectors.

\subsection{Saturation at $\alpha_{\rm sat}\approx 0.01$}

For a more realistic value of $\alpha_{\rm sat}$, we find that the
strain amplitude is about an order of magnitude smaller than for
$\alpha_{\rm sat}$=0.5, but the signal persists for a much longer time
(several months/years for $B\leq 10^{14}$ G).  This is clear from Fig. 
\ref{gwvst0.01}.  This is because of the dependence of $F_m$ on $B^2$ and
$\alpha(t)$.  For smaller $B$ fields and $\alpha_{\rm sat}$, the weaker
magnetic damping allows the r-mode to survive for a longer time, even though
mode growth is slower.  This eventually leads to a {\it faster} spindown for
the star, compared to a higher $\alpha_{\rm sat}$.  To emphasize this
effect, in Fig.~\ref{Pvslogalpha005}, we compare spindown with $\alpha_{\rm
sat}$=0.01 and 0.005, for the same initial stellar configuration and the
same magnetic field.  The crossing of the two curves in that figure
illustrates this effect.  However, we note from comparing Figs. 
\ref{htilde} and \ref{htilde0.01} that $\tilde{h}$ remains relatively
independent of $\alpha_{\rm sat}$, although both the $B=10^{13}$ G and
$B=10^{14}$ G cases shows a flat segment corresponding to the saturated
phase for $\alpha_{\rm sat}=0.01$. Essentially, this is because the
decrease in the peak value of $h(t)$ for smaller $\alpha_{\rm sat}$ is
compensated by the corresponding decrease in $\dot{f}$ (eg., as in
eqn.(\ref{ftsat})).  This implies that we require a larger interval of time
integration to validate the stationary phase approximation for a smaller
$\alpha_{\rm sat}$ (eqn.~\ref{htildeqn}).  Thus, $\tilde{h}$ remains almost
unchanged.

\section{Conclusions} 
\label{conclude} 

We have studied the role of r-modes in spinning a rapidly rotating and
magnetized neutron star down to typical quark deconfinement densities. 
Apart from the usual spindown associated to magnetic braking, the magnetic
damping of r-modes plays an important role in the period evolution of the
star.  We find that, for realistic (small) values of the r-mode saturation
amplitude $\alpha_{\rm sat}$, the time to deconfinement is sped up by the
r-mode instability when magnetic fields are of order $10^{12}$G or less. 
This result reflects the strong dependence of the magnetic damping effect on
$B$ and the evolution of $\alpha$.  Thus, in contrast to the result in
\citep{Lai}, where the r-mode was seen to affect the spindown already for
magnetic fields less than $B\sim 10^{14}$G, the inclusion of the magnetic
damping term leads us to conclude that r-mode driven effects on spindown are
pronounced only for $B\lesssim 10^{12}$G.  We also explored the
gravitational wave signal generated during the spindown phase as the r-mode
first grows then saturates - we follow the signal continuously until the
quark deconfinement threshold is reached.  For realistic values of
$\alpha_{\rm sat}$, the r-mode saturates and leads to a strong signal in
gravitational waves, except in case of very soft equations of state and very
large magnetic field ($10^{15}$ G or more).  There could be a sizeable
fraction of the total rotational kinetic energy emitted as gravitational
waves in this epoch, which can last several years, making it detectable in
upcoming second and third generation detectors.  Therefore, gravitational
waves could be used in this manner to probe the equation of state inside
neutron stars.  As shown here, using r-modes is an alternate way in which
the EoS can be probed through gravitational waves~\footnote{We note that
recently derived empirical upper bounds on the gravitational power radiated by the Crab pulsar constrain the ellipticity of deformed pulsars \citep{Pitkin,Abbott,Virgo} with
implications for the maximum theoretical elasticity of the neutron star
crust \citep{Horowitz}.  On the other hand, in order to similarly constrain
the r-mode amplitude from searches directed at known pulsars, a different
range of frequencies and polarizations must be probed \citep{Owen}, thus the 
limits placed from the Crab pulsar do not have direct bearing on r-modes.}

Gravitational waves are going to open a new window of observation into our
universe. Among the many discoveries that will be made, we anticipate the
exciting prospect that neutron star spindown will reveal signs of the
elusive r-mode instability as well as signatures for the onset of quark
matter in the core. In the appendix we outline a first preliminary estimate of
what the gravitational wave signal from a ``Quark-Nova'' would look like.
It would be interesting to examine this
signal with detailed simulations, especially in cases when the Quark-Nova
occurs very shortly after the neutron star is born in a Supernova - the
signature of this "dual-explosion" in gravitational wave detectors would be
two very different signals coming from the same source but separated in time by a few days to few weeks, depending on the time delay between the two explosions. Such a signature would be unmistakeable in upcoming gravitational wave detectors. 

\appendix {\bf Appendix: Gravitational waves from the quark-hadron phase
transition}

It remains an open question as to what happens to the neutron star when the
deconfinement density is reached. One possibility is that the entire star is
converted to a quark star due to the inherent stability of strange quark
matter \citep{Witten}. If this conversion occurs in an explosive manner,
a``Quark-Nova'' could result \citep{ODD} - what will the gravitational
signal from such an event look like? The conversion involves a two-stage
process - neutrons in the core dissolve into a $u$ and $d$ quark fluid that
is more compact, causing the core to shrink, followed by combustion to
$u,d,s$ quarks through leptonic as well as non-leptonic processes.
\citet{Lin} have studied the first stage with Newtonian hydrodynamics and
found that quadrupolar and quasi-radial modes are excited during the
collapse, leading to gravitational wave emission with an energy output of
~$\sim 10^{51}$ ergs. However, this work applies to a mixed phase of quarks
and nuclear matter and does not consider an explosive transition that begins
with non-premixed fluids. 

\vskip 0.2cm

A full numerical treatment of the second stage, an explosive phase
transition taking into account fluid motion in 3D is a complex task and
beyond the scope of this work.  However, preliminary steps have been taken
in this direction.  \citet{Niebergal10} solved hydrodynamical flow equations
for the combustion of neutron matter to strange quark matter in the laminar
approximation, including weak equilibrating reactions and strange quark
diffusion across the burning front.  The numerical results suggest laminar
speeds of $0.002-0.04$ times the speed of light, much faster than previous
estimates derived using only a reactive-diffusive
description~\citep{Olinto}.  Turbulent combustion has been addressed in a
recent work by \citet{Herzog} who found that the combustion stops short of
converting the entire star to quark matter (the reaction is no longer
exothermic).  This was also the conclusion in \citet{Niebergal10}, though
for a different physical reason~\footnote{\citet{Niebergal10} found that as
the burning front expands and cools, it enters an advection dominated
regime, where the upstream (hadronic) fluid velocity advects the interface
backwards faster than it can progress due to reactions and diffusion. 
Consequently, the interface halts short of the neutron star surface.}. 
Neither of these works continue on to estimate the gravitational signal from
the explosive combustion.  Following the hydrodynamical approach of these
studies, we have estimated the gravitational wave signal $h(t)$ from the
following steps.

\begin{figure}[h!]
\includegraphics[width=0.5\textwidth]{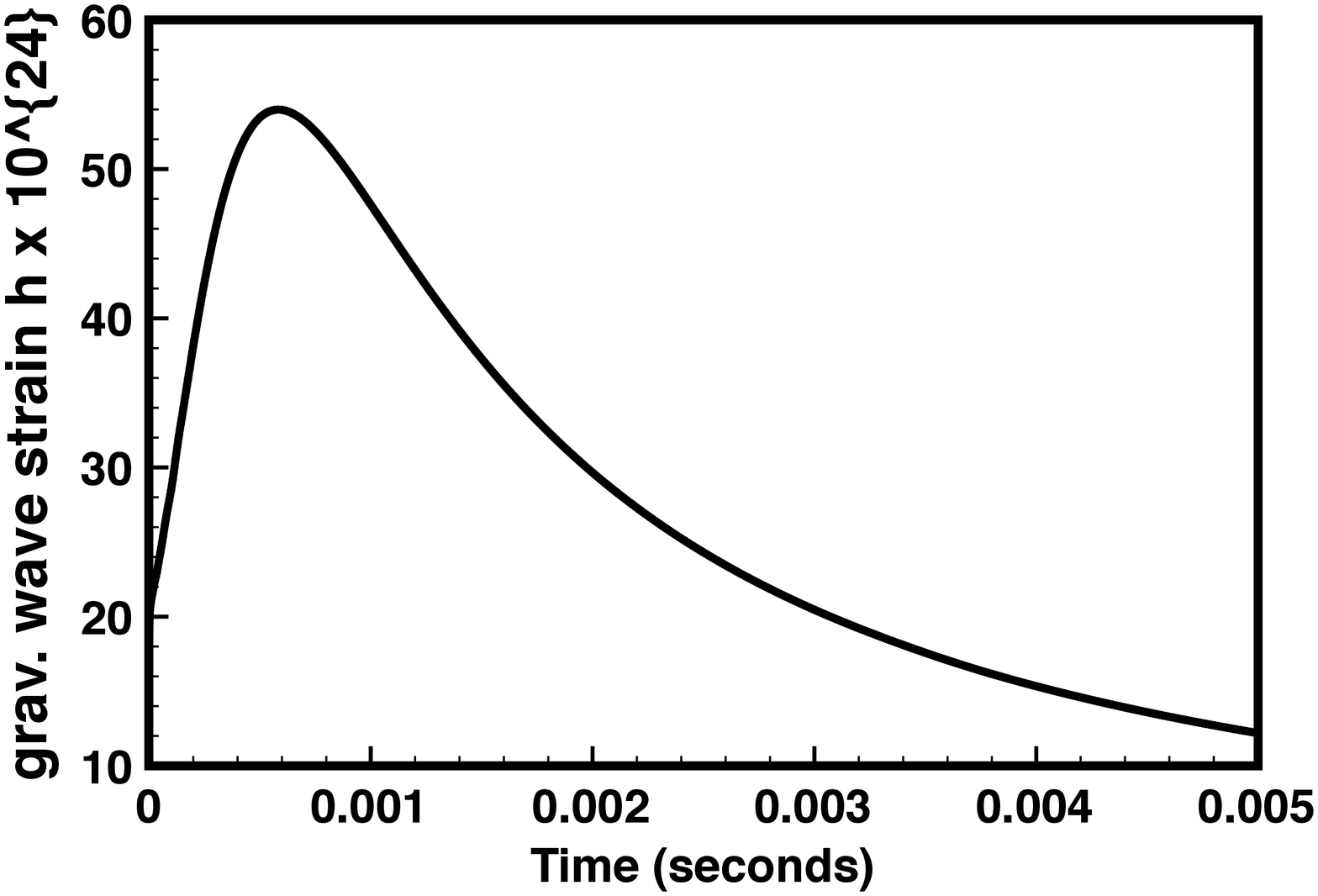}
\includegraphics[width=0.5\textwidth]{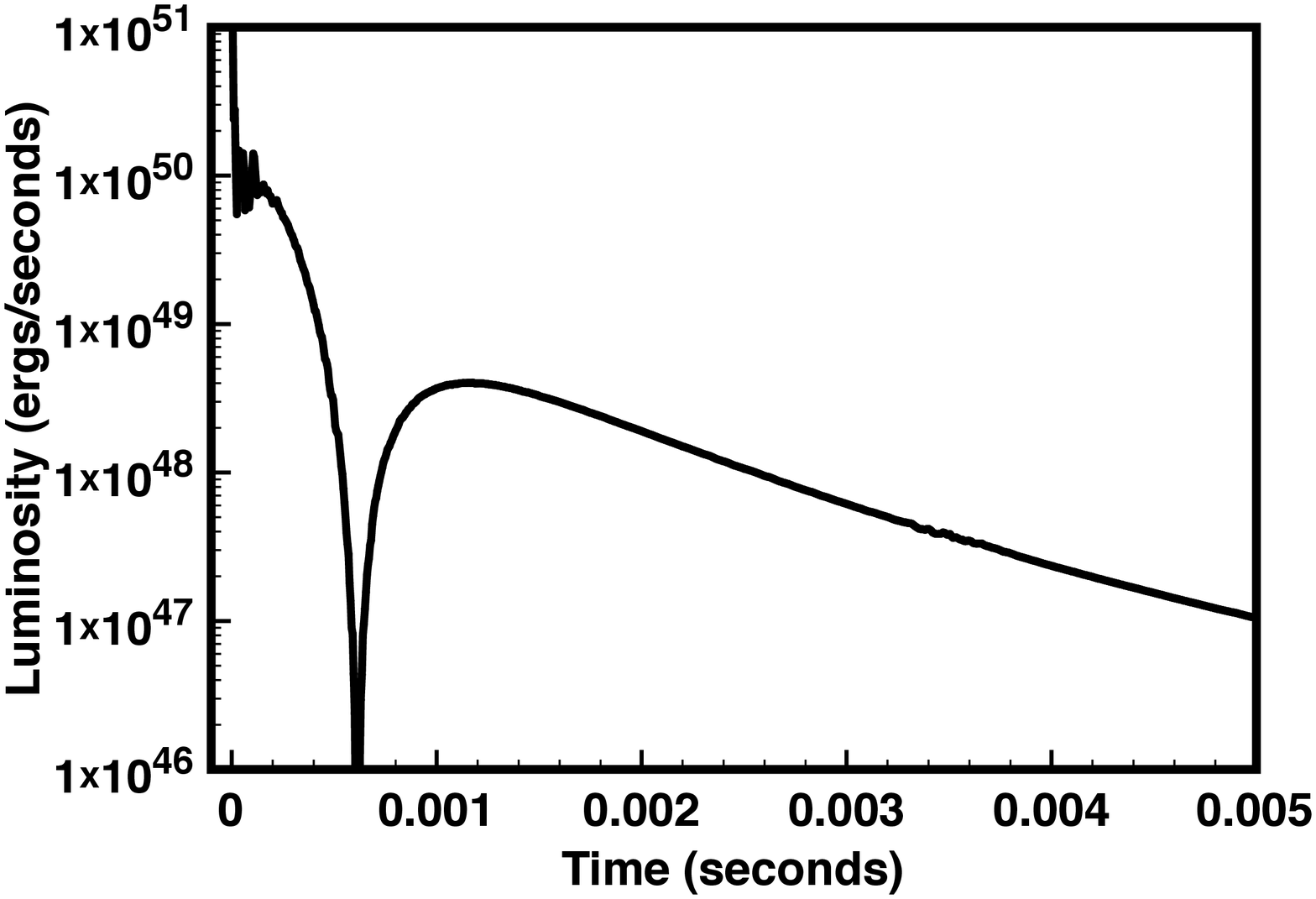}
\caption{Gravitational wave strain $h(t)$ and luminosity $L(t)$ as a
function of time for combustion of neutron matter to strange quark matter following quark deconfinement inside a neutron star.}
\label{QN-signal}
\end{figure}

\begin{itemize}
\item Starting from a mechanically stable configuration where quark matter
constitutes a small fraction of the stellar core, we initiate combustion at
the speeds obtained in \citet{Niebergal10}. The quark fluid, described by a
simple bag model equation of state is subsequently evolved using inviscid
hydrodynamical equations for relativistic fluid flow (relativistic Euler
equations) coupled with Newtonian gravity~\footnote{ Although the
Poisson equation for gravity violates the speed of light (since it is an
elliptic PDE), it is unavoidable unless one uses the full GR.}, assuming
an axisymmetric rotating configuration of the star about the $\hat{z}$-axis.
In the cylindrical coordinates $(r, \phi, \theta)$, where $\phi$ is the
polar angle and $\theta$ is the azimuthal angle, the relevant equations are
given by $\partial_{t} (rU) + \partial_{r}(rF) + \partial_{z}(rG) = S^{'}$,
where
\begin{equation*}
        U = \begin{pmatrix}
                {\rm D} \\ S u_{r} \\ S u_{\theta}\\S u_{z}\\ \tau
        \end{pmatrix}\quad
        F =\begin{pmatrix}
                {\rm D}u_{r} \\ S u_{r}^{2} + p \\ S u_{\theta}u_{r}\\S
u_{z} u_{r}\\ (\tau + p)u_{r}
        \end{pmatrix}\quad
        G = \begin{pmatrix}
                {\rm D}u_{z} \\ S u_{r}u_{z} \\ S u_{\theta} u_{z}\\S
u_{z}^{2} + p\\ (\tau + p)u_{z}
        \end{pmatrix}\quad
        S^{'} = \begin{pmatrix}
                0 \\ Su_{\theta}^{2} + p + rDg_{r} \\ - Su_{\theta}u_{r} \\
r {\rm D} g_{z} \\ rS(u_{r}g_{r} + u_{z}g_{z})
        \end{pmatrix}
\end{equation*}
and where ${\rm D} = \rho \gamma$, $S = {\rm D} h \gamma$, and $\tau =
S-{\rm D}-p$ are introduced solely for the purpose of writing the
relativistic Euler equations in a form analagous to the more familiar
non-relativistic Euler equations.  In the natural units where $c=1$, the
fluid velocity, Lorentz factor, gravitational vector field, and specific
enthalpy are respectively given by $\vec{u}\equiv (u_r,u_{\theta},u_z)$,
$\gamma = 1/\sqrt{1-|\vec{u}|^{2}}$, $\vec{g} = -\nabla \phi$, and $h = 1 +
\epsilon + p/\rho$, where $\epsilon$ is the specific internal energy, $p$ is
the pressure, and $\phi$ is the gravitational potential. In the
non-relativistic limit ($\gamma \rightarrow 1$), we recover the
non-relativistic Euler equations. 

%\item Choosing the quark matter EoS to be given by the Bag model
% ($B^{1/4}$=145 MeV) and the nuclear fluid to be described by a polytrope
% of index $1.7$, the transition density from nucleons to quarks in the
% equilibrium configuration of the star ($M$=1.84$M_{\odot}$, $R$=15km) is
% $?$g/cc. 

\item The quark matter has a density $\rho_{q} = 5 \rho_{n}$
and is described by the Bag model (bag constant $B^{1/4} = 145$ Mev) while the nuclear fluid has density $\rho_{n} = \rho_{\rm sat}=2.5\times 10^{14}{\rm g/cm^3}$ and is described by the perfect caloric EoS with an adiabatic
index of $1.7$.  At $t$=0, we initiate combustion with a density discontinuity $\rho_q-\rho_n$, which is situated at a radial coordinate $r=R/4$ inside the star, and choose an initial burning front speed of $u_r$=0.01$c$, which is in the range of burn velocities obtained in \citet{Niebergal10}. We then evolve the above equations using a weighted average flux relativistic HLL (Harten, Lax, and Van Leer) solver on a cartesian grid with spatial resolution 0.5 km and a timestep determined by a Courant number of $0.3$. 
%We employ the Newton-Raphson method to solve the set of coupled equations
% in $p, \gamma$ to find $\vec{u}, \rho$ at each time step. 
The solver is coupled with the Poisson equation $\nabla^2\phi=4\pi G\rho$.
% which is used to determine the gravitational potential $\phi(\vec{r})$.  

\item Solving for $\vec{u}(r,z,t)$ and $\rho(r,z,t)$, we follow
\citet{Zwerger} to compute the quadrupole wave amplitude $A_{20}^{E2}(t)$
for our axisymmetric configuration, and from there the signal $h(t)$ as
given by eqn.(21) of \citep {Zwerger}.

\begin{equation}
h(t)=\frac{1}{8}\sqrt{\frac{15}{\pi}}{\rm
sin^2\theta}\frac{A_{20}^{E2}(t)}{D}
\end{equation}

We assume a source distance $D=20$ Mpc to plot the maximal signal strength
$h(t)$ and the corresponding luminosity $L(t)$ in Fig.\ref{QN-signal}. The origin of the peak in $h(t)$ is a balance between increasing mass outflow 
and a decreasing density gradient. The luminosity is proportional to the square of $\dot{h}(t)$ (hence the sharp dip at the maximum of $h(t)$) while the total emitted energy obtained from integrating the luminosity curve is $\sim 1.4 \times 10^{48}$ ergs, which is about 5 orders of magnitude smaller than the binding energy of the neutron star ($10^{53}$ ergs) and 2 orders of magnitude less than the energy released in gravitational waves during the first stage of core-collapse, where the energy comes from coupling oscillations to rotational motion~\citep{Lin,Mar}. 

\end{itemize}

Based on this result, the gravitational wave signal for this stage of the phase conversion would be hard to detect in Advanced LIGO, unless the source is Galactic (within few kpc). Most of the energy released in the phase transition is in the form of latent heat and neutrinos. However, our results are only a preliminary estimate, designed to provide a guide to the expected signal from a Quark-Nova. Simulations of
gravitational wave signals from realistic core-collapse supernova models
indicate that convective flows driven by neutrino heating \citep{Muller04}
can drive a strong gravitational wave signal.  The typical time for
neutrinos to diffuse out of the hot quark star \citep[$\approx$ 0.1-1 sec
according to][]{Keranen} is about an order of magnitude larger than the
duration of the signal due to explosive combustion to strange quark matter, therefore, we can expect neutrino heating to be important in our context as well.  We have not included such convective effects in our present simulation, nor the effect of magnetic fields, which can also modify Rayleigh-Taylor instabilities~\citep{Lug02}.

\begin{acknowledgements}
\section*{Acknowledgments}
We are grateful to the anonymous referee who pointed out the importance of magnetic damping on the r-modes. We also thank David Shoemaker and Joel Tohline for comments on the manuscript, Sharon Morsink for help with the RNS code, and Sam Koshy for helpful discussions on gravitational wave theory. J. S. and P. J. would like to thank the hospitality of the department of Physics and Astronomy at the University of Calgary, where part of this work was completed. J. S. is supported, in part, by grant AST-0708551 from the U.S. National Science Foundation and, in part, by grant NNX07AG84G and NNX10AC72G from NASA's ATP program. P. J. and V. C. acknowledge support from from California State University Long Beach and the U. S. Army High Performance Computing Research Center. The research of R. O. is supported by an operating grant from the National Science and Engineering Research Council of Canada (NSERC). 
\end{acknowledgements}

\newpage

\end{document}